\documentclass{aa}

\usepackage{multirow}
\usepackage{graphicx}
\usepackage{txfonts}
\usepackage{lipsum}
\usepackage{subcaption}         % necessary for continued figures, example in section 3
\usepackage{lscape}             % to rotate a single page table, example in appendix.
\usepackage{placeins}           % useful with \FloatBarrier, to keep 
\usepackage{hyperref}
\hypersetup{
	colorlinks=true,       % false: boxed links; true: colored links
	linkcolor=blue,        % color of internal links
	citecolor=blue,        % color of links to bibliography
	filecolor=magenta,     % color of file links
	urlcolor=blue         
}

\usepackage{natbib}
\defcitealias{SPA_DalPonte2025}{DP25}
\defcitealias{Magrini2018}{Magrini et al.\ 2018}
\defcitealias{SalesSilva2022}{Sales-Silva et al.\ 2022}
\defcitealias{Magrini2023}{Magrini et al.\ 2023}
\defcitealias{SPA_Bijavara2025}{Bijavara Seshashayana et al.\ 2025}

\begin{document}

   \title{Stellar Population Astrophysics (SPA) with the TNG}

   \subtitle{Beyond the iron peak: Galactic evolution of molybdenum,  ruthenium and zirconium through open clusters}

%%%%%%%%%%%%%%%%%%%%%%%%%%%%%%%%%%%%%%%%
% Please do not include ORCIDs next to author names.
% Only ORCIDs authenticated by individual authors in EDP Sciences editorial system will be taken into account.
% ORCIDs included here will be removed.
%%%%%%%%%%%%%%%%%%%%%%%%%%%%%%%%%%%%%%%%

   \author{N. Alvarez-Baena\inst{1,2,3}\fnmsep\thanks{ERASMUS+ Master in Astrophysics and Space Science (MASS) Alumni.} \and V. D'Orazi\inst{1,2} \and C. Sneden\inst{4} \and A. Psaltis\inst{5} \and F. Th\'{e}venin\inst{3} \and A. Bragaglia\inst{6}  \and T. Mishenina\inst{7} \and S. Bijavara Seshashayana\inst{8,9} \and R. Carrera\inst{6} \and L. Casamiquela\inst{10} \and M. Dal Ponte\inst{11} \and H. Jönsson\inst{8} \and A. Nunnari\inst{1,2} \and M. Pignatari \inst{12,13,14,15} 
   }
   \institute{Department of Physics, University of Rome Tor Vergata, via della Ricerca Scientifica 1, 00133 Rome, Italy\\
             \email{natalia.alvarezbaena@students.uniroma2.eu}
            \and INAF - Osservatorio Astronomico di Roma, via Frascati 33, 00178, Monte Porzio Catone, Italy
            \and Université C\^{o}te d’Azur, Observatoire de la
Côte d’Azur, CNRS, Laboratoire Lagrange, France
\and Department of Astronomy and McDonald Observatory, The University of Texas, Austin, TX 78712, USA
\and Department of Astronomy \& Physics, Saint Mary’s University, Halifax, NS B3H 3C3, Canada
\and INAF – Osservatorio di Astrofisica e Scienza dello Spazio di Bologna, via Piero Gobetti 93/3, 40129 Bologna, Italy
\and Astronomical Observatory, Odesa National University, 1b, Marazlievska str., 65014 Odesa, Ukraine
\and Materials Science and Applied Mathematics, Malmö University, SE-205 06 Malmö, Sweden
\and Nordic Optical Telescope, Rambla José Ana Fernández Pérez 7, ES-38711 Breña Baja, Spain
\and LIRA, Observatoire de Paris, Université PSL, Sorbonne Université, Université Paris Cité, CY Cergy Paris Université, CNRS,
92190 Meudon, France
\and INAF-Osservatorio Astronomico di Padova, vicolo dell’Osservatorio 5, 35122 Padova, Italy
\and Konkoly Observatory, HUN-REN, Konkoly Thege Miklos ut 15-17, H-1121 Budapest, Hungary
\and MTA Centre of Excellence, Budapest, Konkoly Thege Miklós út 15-17 H-1121, Hungary
\and University of Bayreuth, BGI, Universitätsstraße 30, 95447 Bayreuth, Germany
\and NuGrid Collaboration, \url{http://www.nugridstars.org}
}

   \date{Received September 30, 20XX}

% \abstract{}{}{}{}{}
% 5 {} token are mandatory
 
  \abstract
   {Open clusters serve as key laboratories for studying the chemical evolution of the Milky Way. In particular, leveraging these coeval groups of stars to study elements beyond the iron peak provides insights into heavy-element nucleosynthesis, whose origins remain uncertain.}
   {This work aims to determine the Mo, Ru, and Zr abundances for 81 stars across 30 open clusters. We compare these new chemical patterns with those of key reference elements (Y, Sr, Eu), which were previously derived by our group. This comparison seeks to disentangle the distinct contributions from various nucleosynthetic processes.} 
   {Using the high-resolution spectra acquired within the Stellar Population Astrophysics programme using the HARPS-N echelle spectrograph at the Telescopio Nazionale Galileo, we estimate chemical abundances through spectral synthesis with \texttt{TSFitPy} under local thermodynamic equilibrium (LTE) conditions.}
   {This study provides the first Ru abundances for a large open cluster sample, alongside new measurements for Mo and Zr. We observe tight correlations between Mo, Ru, and other s-process elements (Sr, Y, Zr), with slopes near unity, suggesting common enrichment timescales. Conversely, relations with the r-process element Eu show significant deviations, implying additional nucleosynthetic contributions. Crucially, chemical planes involving [Mo/Ru] and [s/Mo] ratios suggest systematic offsets when compared to theoretical s-process predictions.}
  % conclusions heading (optional)
  {}


   \maketitle

    \nolinenumbers
%%%%%%%%%%%%%%%%%%%%%%%%%%%%%%%%%%%%%%%%%%%%%%%%%%%%%%%%%%%%%%
\section{Introduction}

Spanning the complete lifespan of the Galactic thin disc \citep{Bossini2019}, open clusters (OCs) provide a unique window into the chemo-dynamical evolution of the Milky Way \citep{Cantat-Gaudin2018, Spina2022}. Their importance in this field has been demonstrated across several recent works \citep[e.g.,][hereafter \citetalias{SPA_DalPonte2025}]{Carrera2019M67, Alvarez-Baena2024, SPA_Zhang2022, Carbajo-Hijarrubia2024, SPA_DalPonte2025}. While Gaia \citep{Gaia_2016Prusti} provides unprecedented astrometric data for determining stellar distances and kinematics, its Radial Velocity Spectrometer (RVS) is constrained by a relatively modest spectral resolution and limited precision for detailed elemental abundances (see e.g., \citealt{Carrera2022}). Consequently, several ground-based spectroscopic surveys have been established to complement Gaia’s reach. High and medium-resolution projects such as the Gaia-ESO Survey \citep{Gilmore2012,Randich2022}, APOGEE \citep{Majewski2017}, and GALAH \citep{DeSilva2015}, alongside forthcoming projects like WEAVE \citep{Jin2024}, 4MOST \citep{Dejong2019}, and MOONS \citep{Cirasuolo2014}, provide the necessary depth. These are further augmented by high-resolution, targeted surveys for OCs, including OCCASO \citep[Open Clusters Chemical Abundances
from Spanish Observatories,][]{Casamiquela2016}, OSTTA \citep[One Star to Tag Them All,][]{Carrera2022}, and SPA \citep[Stellar Population Astrophysics,][]{SPA_Origlia19}. Together, these homogeneous datasets allow for a high-precision mapping of chemical abundance trends across the Galactic disc, revealing the Milky Way's enrichment history \citep{Spina2022}.

Within this framework, neutron-capture elements serve as critical tracers of both nucleosynthetic processes and the broader chemical evolution of the Galaxy. While several investigations have successfully characterised neutron-capture species in OCs \citepalias[e.g.,][]{Magrini2018, SalesSilva2022, Magrini2023,SPA_DalPonte2025,SPA_Bijavara2025} including molybdenum (Mo, $Z=42$; \citealt{Overbeek2016, Mishenina2020, Casamiquela2021, VanderSwaelmen2023}), ruthenium (Ru, $Z=44$) remains largely unobserved in statistically significant OC samples.

A joint analysis of Mo and Ru is essential, as their different fractional contributions from multiple nucleosynthetic processes (see Sect.~\ref{sec:nucleosynth}) provide a powerful diagnostic of their origin when considered together. While previous studies have investigated Mo and Ru in the Galactic disc, halo, bulge and field stars \citep[see e.g.,][]{Hansen2014,  Spite2018, Mishenina2019, Mishenina2026, Forsberg2022}, their behaviour in OCs remains largely unexplored. In this work, we exploit the high-resolution spectra of the SPA programme to derive, for the first time, Ru abundances for a large sample of OCs, alongside their Mo measurements, providing new insights into the chemical evolution of these neutron-capture elements in the Galactic disc. Additionally, we derived abundances for the s-process element zirconium (Zr, Z\,=\,40) in our sample, and complement our analysis with the results obtained by \citetalias{SPA_DalPonte2025}, for the s-process elements strontium (Sr, Z\,=\,38) and yttrium (Y, Z\,=\,39), as well as the r-process element europium (Eu, Z\,=\,63).

This paper is organised as follows. We present a general overview about the nucleosynthetic processes involved in the production of Mo and Ru in Sect.~\ref{sec:nucleosynth}.  In Sect.~\ref{sec:observations}, we describe the observations and the selected sample. The determination of Mo, Ru, and Zr abundances are discussed in Sect.~\ref{sec:specanalyses}. Our results, together with a comparison to literature data and current models, are discussed in Sect.~\ref{sec:results}. Finally, Sect.~\ref{sec:conclu} summarises our main findings and conclusions.

\section{Production pathways of Mo and Ru} \label{sec:nucleosynth}

While the synthesis of lighter nuclei, up to and including iron (Fe, Z\,=\,26), occurs through thermonuclear stellar fusion reactions, elements beyond Fe proceed through different pathways, among which neutron-capture is a major contributor \citep{Karakas2014}. This process unfolds in two stages: first, a neutron is captured by a seed nucleus, forming a heavier isotope; then, if the resulting nucleus is unstable, it undergoes a $\beta^-$--decay ($n \rightarrow p + e^- + \Bar{\nu_{e}}$). The subsequent differentiation between the slow (s-) and rapid (r-) processes is given by comparing the rate of neutron-captures (which depends on the neutron flux density) and the $\beta^-$--decay timescale \citep{Burbidge1957, Cowan2021}.

The s-process consists of three different components \citep[see][and references therein]{Busso1999, Bisterzo2014, Mishenina2019}. The weak component that contributes to the s-isotopes from Fe to Sr, takes place in massive stars due to the $^{22}$Ne($\alpha$, n)$^{25}$Mg reaction,  which is activated at the end of the convective He-burning core and in the subsequent convective C-burning shell. The main component contributing for the s-nuclei from Sr to Pb, is produced by low- to intermediate-mass Asymptotic Giant Branch (AGB) stars, where neutrons are released by the  $^{13}$C($\alpha$, n)$^{16}$O reaction in the radiative $^{13}$C-pocket, formed
right after the third dredge-up event and the $^{22}$Ne($\alpha$, n)$^{25}$Mg reaction that is mildly activated during the convective thermal pulses. The strong component is expected to provide at least 50$\%$ of $^{208}$Pb, which is the most abundant nucleus produced in low metallicity AGB stars at [Fe/H] $\sim$ --1 \citep[e.g.,][]{Gallino1998}.

Furthermore, the r-process is responsible for producing about half of the heavy elements, yet its dominant astrophysical site remains uncertain \citep{Cowan2021}. The two main candidate sources are core-collapse supernovae (CCSNe) and neutron star - neutron star (NS-NS) mergers  \citep{Cote2018}. The weak component of the r-process is associated with the neutrino-driven winds from CCSNe and electron-capture supernovae \citep{Wanajo2011, ArconesThielemann2013}. The main component is associated with the neutron-rich matter ejected by NS-NS mergers and neutron star - black hole mergers \citep[see][and the references therein]{Mishenina2019}, as well as magneto-rotational supernovae (MR-SNe; \citealt{Reichert2021}).

An additional intermediate (i-) process was proposed by \citet{Cowan1977} where the neutrons are mainly produced by the reaction $^{13}$C($\alpha$, n)$^{16}$O. As explained by \citet{Choplin2021}, the i-process is triggered when hydrogen is mixed in a convective He-burning zone. Protons are transported down in the convective He-burning zone and are burned by the $^{12}$C(p, $\gamma$)$^{13}$N reaction. The $^{13}$N decays in $\sim$ 10 minutes into $^{13}$C, being followed by the reaction $^{13}$C($\alpha$, n)$^{16}$O that operates mainly at the bottom of the convective helium-burning zone.

Besides the neutron-capture, the proton-capture (or p-process) is responsible for the production of stable proton-rich nuclides that cannot be formed through neutron-capture processes \citep[see][and references therein]{Arnould2003,Rauscher2013}. The $\gamma$-process is the most established scenario for the production of the p-nuclei, which are synthesised via different photo-disintegration paths starting from heavier nuclei \citep{Pignatari2016}. It proceeds through reactions that can lead to the formation of p-nuclei in two ways: either by destroying pre-existing neutron-richer isotopes through sequences of ($\gamma$, n) reaction -- which dominates the most stable nuclei -- or by initiating reaction chains from heavier, unstable nuclei, where ($\gamma$, p) or ($\gamma$, $\alpha$) reactions occur, followed by a series of $\beta^-$--decays that gradually produce proton-rich isotopes. The favoured scenario for the p-process is associated with the explosion of the Ne/O-rich layers of a massive star during its final core collapse as a Type II supernova (SNII). However, it fails to produce the required amounts of p-nuclei in several mass ranges \citep{Rauscher2013}. For this reason, other complementary mechanisms have been proposed, e.g. Type Ia Supernovae, He-accreting CO white dwarfs of sub-Chandrasekhar mass and proton-rich components of neutrino-driven winds \citep[see][and all the references therein]{Mishenina2019}.

Leveraging the well-determined ages of OCs, studies have demonstrated that s-process abundances with respect to iron ([s/Fe]) increase in younger stellar populations-- a trend first identified by \citet{Dorazi2009} for barium (Ba, Z = 56) and further explored later on (see \citealt{Magrini2022} for a review). However, we notice that Ba has been proven to be very uncertain at a young age and lanthanum (La, Z = 57) should always be preferred as an s-process tracer \citep{dorazi2022}.

Both Mo and Ru consist of seven stable isotopes each, produced through different nucleosynthetic channels, and measuring their abundances offers key insights into the chemical evolution of the Galaxy \citep{Mishenina2026}. The $^{92,94}$Mo are p-only nuclei \citep{Rauscher2013,Pignatari2016}, while $^{96}$Mo is a pure s-process isotope \citep{Kapeller2011, Bisterzo2014} and $^{100}$Mo is an r-only isotope \citep{Travaglio2004}. Additionally, the i-process could produce efficiently $^{95}$Mo and $^{97}$Mo \citep{Coteip2018}. 
Similarly, $^{96,98}$Ru are p-only nuclei \citep{Rauscher2013,Pignatari2016}, while $^{100}$Ru is a a pure s-isotope \citep{ Bisterzo2014} and $^{104}$Ru is an r-only isotope \citep{Sneden2008, Prantzos2020}. The different contributions for each isotope are presented in Table~\ref{table:mo_ru_isotopes_full}. 
While not considered in this classification, neutrino-driven ejecta from core-collapse supernovae have also been proposed to explain Mo and Ru observations in a number of metal-poor stars and for a possible impact on the GCE of these elements \citep[e.g.,][]{Montes2007, farouqi:10, arcones:11, Wanajo2011, bliss:18, Psalti2024}. In this case, the relative contribution to the different isotopes may vary greatly depending on the properties of the components considered.

The even isotopes ($^{92,94,96,98,100}$Mo) have nuclear spin $I=0$ and therefore produce no hyperfine splitting, while the odd isotopes ($^{95,97}$Mo), with spin $I=1/2$, should exhibit such structure. However, \citet{Grace1934} showed that the hyperfine components of the odd isotopes are extremely weak and become masked by the much stronger contributions of the even isotopes. In addition, small isotope shifts overlap with the expected hyperfine separations, so the observed \ion{Mo}{i} lines appear only as broadened or asymmetric profiles in their lab spectra, preventing a clean measurement of hyperfine splitting. For this reason, the Mo isotopic ratios cannot be determined, but rather the Mo abundance as a whole, which enables the tracing of major contributions from various nucleosynthesis processes \citep{Hansen2014,Forsberg2022}. An analogous situation is expected for Ru \citep{Murakawa1955}.

\begin{table*}
    \caption{Solar abundances of Mo and Ru isotopes and their contribution by mass fraction as reported by \citet{Sneden2008}.}
    \centering
    \begin{tabular}{cccccc}
        \hline\hline
        Isotope & Solar Abundance (\%)\tablefootmark{a} & Processes & s-process\tablefootmark{h} & r-process\tablefootmark{h} & p-process \\
        \hline\\
        $^{92}$Mo  & 14.525 & p-only\tablefootmark{bcd}            & $-$    & $-$    & 1.000 \\
        $^{94}$Mo  & 9.151  & p-only\tablefootmark{bcd}            & $-$    & $-$    & 1.000 \\
        $^{95}$Mo  & 15.838 & (r-, s-)\tablefootmark{bhi} and i-\tablefootmark{g} & 0.470 & 0.530 & $-$ \\
        $^{96}$Mo  & 16.672 & s-only\tablefootmark{befhi}         & 1.000 & $-$    & $-$ \\
        $^{97}$Mo  & 9.599  & (r-, s-)\tablefootmark{bhi} and i-\tablefootmark{g} & 0.642 & 0.358 & $-$ \\
        $^{98}$Mo  & 24.391 & (r-, s-)\tablefootmark{bhi} & 0.847 & 0.153 & $-$ \\
        $^{100}$Mo & 9.824  & r-only\tablefootmark{bh} or (r-, s-)\tablefootmark{i} & $-$    & 1.000 & $-$ \\
        \hline\\
        $^{96}$Ru & 5.540  &  p-only\tablefootmark{bcd} & $-$  & $-$ & 1.000  \\
        $^{98}$Ru & 1.870  & p-only\tablefootmark{bcd} &  $-$ & $-$ & 1.000 \\
        $^{99}$Ru &  12.760 & (r-, s-)\tablefootmark{bi} or s-only \tablefootmark{h}& 1.000   & $-$  &  $-$ \\
        $^{100}$Ru & 12.600  &   s-only\tablefootmark{bfhi} & 1.000  & $-$ & $-$ \\
        $^{101}$Ru & 17.060  & (r-, s-)\tablefootmark{bhi}   &  0.158 & 0.842 & $-$ \\
        $^{102}$Ru &  31.550 & (r-, s-)\tablefootmark{bhi}  &  0.444 & 0.556 & $-$ \\
        $^{104}$Ru & 18.620  & r-only\tablefootmark{bhi}  & $-$  & 1.000 & $-$ \\
        \hline
    \end{tabular}
    \tablefoot{
    \tablefoottext{a}{\citet{Asplund2009}} 
    \tablefoottext{b}{\citet{AndersGrevesse1989}} 
    \tablefoottext{c}{\citet{Rauscher2013}} 
    \tablefoottext{d}{\citet{Pignatari2016}} 
    \tablefoottext{e}{\citet{Kapeller2011}} 
    \tablefoottext{f}{\citet{Bisterzo2014}} 
    \tablefoottext{g}{\citet{Cote2018}} 
    \tablefoottext{h}{\citet{Sneden2008}} 
    \tablefoottext{i}{\citet{Prantzos2020}}.
    }
    \label{table:mo_ru_isotopes_full}
\end{table*}

\section{Observations and the cluster sample} \label{sec:observations}

    The SPA Programme exploits the combination of the High Accuracy Radial velocity Planet Searcher for the Northern Hemisphere (HARPS-N) and GIANO-B echelle spectrographs at the 3.5-metre Telescopio Nazionale Galileo (TNG) at the Observatorio del Roque de los Muchachos on La Palma (Canary Islands, Spain). It aims at obtaining high-quality spectra of more than 500 stars in the Milky Way thin disc and associated star clusters at different Galactocentric distances, including the poorly explored inner disc. 
    
    The SPA project includes both young OCs, such as ASCC 123 \citep{SPA_Frasca2019}, Praesepe \citep{SPA_DOrazi2020}, and Stock 2 \citep{AlonsoSantiago2021}, as well as older ones, such as Collinder\,350, Gulliver\,51, NGC\,7044, and Ruprecht\,171 \citep{SPA_Casali2020}. Additionally, SPA  exploits the NIR instrument GIANO-B to study helium \citep{SPA_Jian2024}, phosphorus \citep{SPA_Jian2026}, fluorine \citep{SPA_Bijivara2024}, and many other elements, to be compared with what we obtain from HARPS-N \citep{SPA_Bijavara2025}.
    In particular, \citetalias{SPA_DalPonte2025} estimated the non-local thermodynamic equilibrium (NLTE) atmospheric parameters and abundances of giant stars in 33 OCs, from the HARPS-N spectra. Their results  will serve as the starting point for this study.
    
    For this work, we restrict ourselves to the analysis of the HARPS-N optical spectra with a resolution of R $\approx$ 115 000 and a spectral coverage $\lambda$ $\approx$ 3800 - 6900 \r{A} \citep{Cosentino2012}. The observations used for this study were obtained between July 2018 and April 2023. 

    Our sample consisted initially of 93 evolved stars on the red giant branch (RGB) and the red clump (RC) across 33 different OCs (see Fig.~\ref{fig:cmds}) selected from high-probability OC members based on \citet{Cantat-Gaudin2018, Cantat-Gaudin2020}. These numbers differ from \citetalias{SPA_DalPonte2025} as we discarded the two stars in Theia\,1214 (included as members of NGC\,752 in \citetalias{SPA_DalPonte2025}) since our analysis revealed large discrepancies in their chemical abundances both between the two stars and with respect to NGC\,752. This result is in agreement with \citet{SPA_CNO_Curjuric2026}, who performed a dedicated analysis of these two stars, concluding that they are most likely field stars. An additional constraint in effective temperature (T$_{\text{eff}}$) was applied: stars with T$_{\text{eff}}$ below 4300 K were discarded as we were unable to properly estimate C and N for them as their resulting spectral fits showed an over-depression of the continuum that we believe to be related to the presence of TiO molecular bands. A more detailed discussion of this limitation will be presented in a dedicated paper (Alvarez-Baena, in prep.). After this cut, the final sample consists of 81 stars in 30 OCs as Gulliver\,18, Gulliver\,24 and NGC\,7044 were entirely removed because all of their members fell below this threshold. The final number of stars in each OC, along with the general properties of the clusters, is presented in Table~\ref{table:OC_properties}. 

    \begin{table*}
\caption {Summary of the OC analysed in this work. For each cluster, we list the number of stars studied in this work (N) and in \citetalias{SPA_DalPonte2025} (SPA members). The OC properties (Age, A$_{\text{v}}$, distance, R$_{\text{GC}}$) are from \citet{Cantat-Gaudin2020}. \newline}
\label{table:OC_properties} 
\centering
\begin{tabular}{llcccccc}
\hline\hline             
Cluster & Other names &  SPA members\tablefootmark{a}  & N & Age & A$_{\text{v}}$  & Distance  & R$_{\text{GC}}$ \\
 &  &   & & (Gyr) &  (mag) & (pc) &  (pc)  \\
\hline
Alessi 1 & Casado-Alessi & 4 & 4 & 1.45 & 0.08 & 689 & 8726 \\
Alessi 161 & UBC 3  & 1 & 1 & 0.13 & 0.98 & 1704 & 7223 \\
Alessi-Teutsch 11 & ASCC 112, Alessi 46 & 1 & 1 & 0.14 & 0.37 & 634 & 8335 \\
Basel 11b & FSR 877 & 3 & 3 & 0.23 & 1.56 & 1793 & 10121 \\
COIN-Gaia 30 & $-$ & 1 & 1 & 0.26 & 1.25 & 767 & 8804 \\
Collinder 350 & $-$ & 2 & 1 & 0.59 & 0.52 & 371 & 8021 \\
Collinder 463 & $-$ & 2 & 2 & 0.11 & 0.79 & 849 & 8874 \\
Gulliver 51 & $-$ & 1 & 1 & 0.36 & 1.42 & 1536 & 9410 \\
IC 4756 &  Collinder386, Melotte 210 & 12 & 12 & 1.29 & 0.29 & 506 & 7938 \\
LP 1800 & UBC 170, FoF 1800  & 3 & 3 & 0.33 & 1.25 & 1392 & 8809 \\
NGC 2437 & M 46, Melotte 75 & 5 & 5 & 0.30 & 0.73 & 1511 & 9345 \\
NGC 2509 & Melotte 81, Collinder 171 & 1 & 1 & 1.51 & 0.23 & 2495 & 9887 \\
NGC 2548 & M 48, Melotte 85 & 3 & 3 & 0.40 & 0.15 & 772 & 8857 \\
NGC 2632 & M 44, Praesepe & 2 & 2 & 0.68 & 0.00 & 183 & 8479 \\
NGC 2682 & M 67 & 2 & 2 & 4.27 & 0.07 & 889 & 8964 \\
NGC 6800 & $-$ & 1 & 1 & 0.41 & 0.83 & 1012 & 7868 \\
NGC 6991 & $-$ & 5 & 5 & 1.55 & 0.20 & 577 & 8333 \\
NGC 7086 & Collinder 437  & 2 & 2 & 0.19 & 1.81 & 1677 & 8632 \\
NGC 7209 & Melotte 238, Collinder 444  & 2 & 1 & 0.43 & 0.53 & 1154 & 8525 \\
NGC 752 & Melotte 12, Theia 1214 & 5 & 3 & 1.17 & 0.07 & 483 & 8669 \\
Ruprecht 171 & $-$ & 7 & 4 & 2.75 & 0.68 & 1522 & 6895 \\
Stock 2 & $-$ & 8 & 8 & 0.40 & 0.50 & 399 & 8619 \\
Tombaugh 5 & $-$ & 3 & 3 & 0.19 & 2.07 & 1706 & 9768 \\
UBC 141 & $-$ & 1 & 1 & 2.09 & 0.51 & 1315 & 8176 \\
UBC 169 & $-$ & 2 & 1 & 0.30 & 0.26 & 1347 & 8749 \\
UBC 194 & $-$ & 1 & 1 & 0.23 & 0.48 & 1361 & 9394 \\
UBC 577 & Alessi 191 & 4 & 4 & 2.75 & 0.00 & 1122 & 7718 \\
UBC 60 & COIN-Gaia 11 & 1 & 1 & 0.79 & 1.25 & 669 & 8977 \\
UPK 219 & $-$ & 1 & 1 & 0.15 & 1.20 & 873 & 8735 \\
UPK 84 & UBC 131, Alessi 116 & 3 & 3 & 1.00 & 0.31 & 944 & 7980\\
\hline
\end{tabular}
\tablefoot{\tablefoottext{a}{Stars studied in each OC by \citetalias{SPA_DalPonte2025}. }}
\end{table*}

    For this study, we used the stellar parameters and chemical abundances of Sr, Y and Eu derived by \citetalias{SPA_DalPonte2025}. For the derivation of the stellar parameters, \citetalias{SPA_DalPonte2025} adopted the NLTE Optimization Tool Utilized for the derivation of atmospheric Stellar parameters (\texttt{LOTUS}) code\footnote{\url{https://github.com/Li-Yangyang/LOTUS}} \citep{Li2023}, which derives T$_{\text{eff}}$, surface gravity ($\log g$), metallicity ([Fe/H]) and microturbulence ($v_{\text{mic}}$) from \ion{Fe}{i} and \ion{Fe}{ii} EWs using generalised curves of growth in LTE and NLTE. Parameters are obtained via global minimisation, with uncertainties estimated through a Markov chain Monte Carlo (MCMC) approach. The elemental abundances were obtained using NLTE synthesis \citepalias[for details see][]{SPA_DalPonte2025}. Since macroturbulence ($v_{\text{mac}}$) values were not explicitly reported by \citetalias{SPA_DalPonte2025}, we derived them by averaging the results from their successful Ti line fits. For two exceptions, UBC~169~star~\#2 and UBC~194~star~\#1, the $v_{\text{mac}}$ values were instead calculated based on Ni line estimations. The adopted values are summarised in Table~\ref{table:stellarparams}.

\section{Spectroscopic analysis} 
\label{sec:specanalyses}

We used the open-source \texttt{PYTHON} wrapper \texttt{TSFitPy}\footnote{\url{https://github.com/TSFitPy-developers/TSFitPy}} \citep{Gerber2023} of the \texttt{Turbospectrum 2020} code \citep{Alvarez1998,Plez2012}\footnote{\url{https://github.com/bertrandplez/Turbospectrum\_NLTE}}. This tool allows the computation of synthetic stellar spectra with NLTE for multiple chemical species at once by fitting the normalised synthetic spectra from \texttt{Turbospectrum} by $\chi^2$ minimisation using the Nelder–Mead algorithm \citep{Gerber2023}. The tool was further optimised and had the addition of some extra features as described in \citet{Storm2023}, including the microturbulence optimisation as described in \citet{DOrazi_Galah}. The assumed solar abundances from \citet{Asplund2009} and \citet{Magg2022} are: A(Fe)$_\odot$ = 7.50, A(Sr)$_\odot$ = 2.87, A(Y)$_\odot$ = 2.21, A(Zr) = 2.58, A(Mo)$_\odot$ = 1.88, A(Ru)$_\odot$ = 1.75, and A(Eu)$_\odot$ = 0.52.

Our initial analysis focused on the \ion{Mo}{i} lines at 5506 and 5533 \r{A} \citep{Mishenina2019, Mishenina2026} and the 6030 \r{A} line \citep{Forsberg2022}. Following the \textit{Gaia}-ESO line list classification \citep{Heiter2021_GESline}, the 6030 \r{A} feature is categorised as a \textit{Yes/Yes} line, indicating both a high-quality oscillator strength ($\log~gf$) value and an unblended profile. This classification system provides usage recommendations for spectral lines based on their performance in solar and Arcturus benchmark spectra. In contrast, the 5506 \r{A} and 5533 \r{A} lines are classified as \textit{Yes/Undecided} and \textit{Yes/No}, respectively. Upon inspecting our sample, we found the \ion{Mo}{i} lines at 5506 and 5533 \AA\ to be severely affected, even at very high resolution, by the presence of strong nearby \ion{Fe}{i} features. This rendered any resulting fits unreliable; consequently, these two lines were excluded from our analysis. The 6030 \AA\ line proved to be too weak for reliable measurement in the majority of our spectra. Relying solely on the 6030 \AA\ feature would have reduced our effective sample size to approximately one-third of its original total. Due to the limitations found for the 6030 \AA\ line, we tested several other lines, e.g. 5751.408 (\textit{Yes/Yes}) and  5858.266 ($Yes/Undecided$). We found that a good alternative was the \ion{Mo}{i} line at 5570.444 \r{A} (\textit{Yes/Undecided}) previously used by \citet{Roriz2021} for the Mo estimation in Ba stars and by \citet{Overbeek2016} for Mo determination in OCs. This line was successfully measured for the majority of our sample.

\begin{table}[h!]
\caption{Atomic data for the spectral lines analysed in this work. This data comes from the line list of the \textit{Gaia}-ESO Survey  \citep{Heiter2021_GESline} updated by \citet{Magg2022}. The atomic parameters for \ion{Mo}{i}, \ion{Ru}{i}, and \ion{Zr}{i} were originally reported by \citet{WBb}, \citet{WSL}, and \citet{BGHL}, respectively.}                 % title of Table
\label{table:atomicparams}    % is used to refer this table in the text
\centering                        % used for centering table
\begin{tabular}{c c c c}      % centered columns (4 columns)
\hline\hline               % inserts double horizontal lines
Species & Wavelength in air & $\log~gf$ & $\chi^{\text{low}}_{\text{exc}} $  \\
      & (\r{A}) &  (dex) &  (eV) \vspace{1mm}\\
    \hline
    \ion{Mo}{i} & 5570.444 & -0.337 & 1.335\\
    \ion{Mo}{i} & 6030.644 & -0.523 & 1.531 \\
    \ion{Ru}{i} & 4869.153 & -0.830 & 0.928\\
    \ion{Zr}{i} & 6127.440 & -1.060 & 0.154 \\
    \ion{Zr}{i} & 6134.5500 & -1.280 & 0.000 \\
\hline                                  %inserts single line
\end{tabular}
\end{table}

The \ion{Mo}{i} lines at 5570 and 6030 \r{A} are both affected by CN molecular blends, with the contribution being particularly significant for the 5570 \r{A} feature. Although \texttt{TSFitPy} defaults to solar abundances, the giant stars in our sample are expected to exhibit nitrogen enhancement and carbon depletion \citep{Karakas2014}. To accurately model the CN blending, we independently determined C and N abundances for each star using the $^{12}$C$_2$ Swan (0,1) band with band head 5635 \r{A} and the $^{12}$C$^{14}$N molecular bands in the spectral range 6470-6490 \r{A}. We adopt [O/Fe] = 0.0 dex because no telluric spectra were available. For more than 90\% of our sample, the \ion{O}{i} line at 6300.3040 \r{A} is contaminated by telluric absorption features, a well-known limitation of this diagnostic \citep[e.g.,][]{VanPlas08_Oline, Jonsson17_Oline}. This assumption is supported by the results of \citetalias{SPA_DalPonte2025}, who found nearly solar metallicities for 33 OCs consistent with expectations for stellar populations in the Milky Way thin disc \citep{gaiadr3_recioblanco23}.

\begin{figure*}[ht!]
\centering
\includegraphics[width=\linewidth]{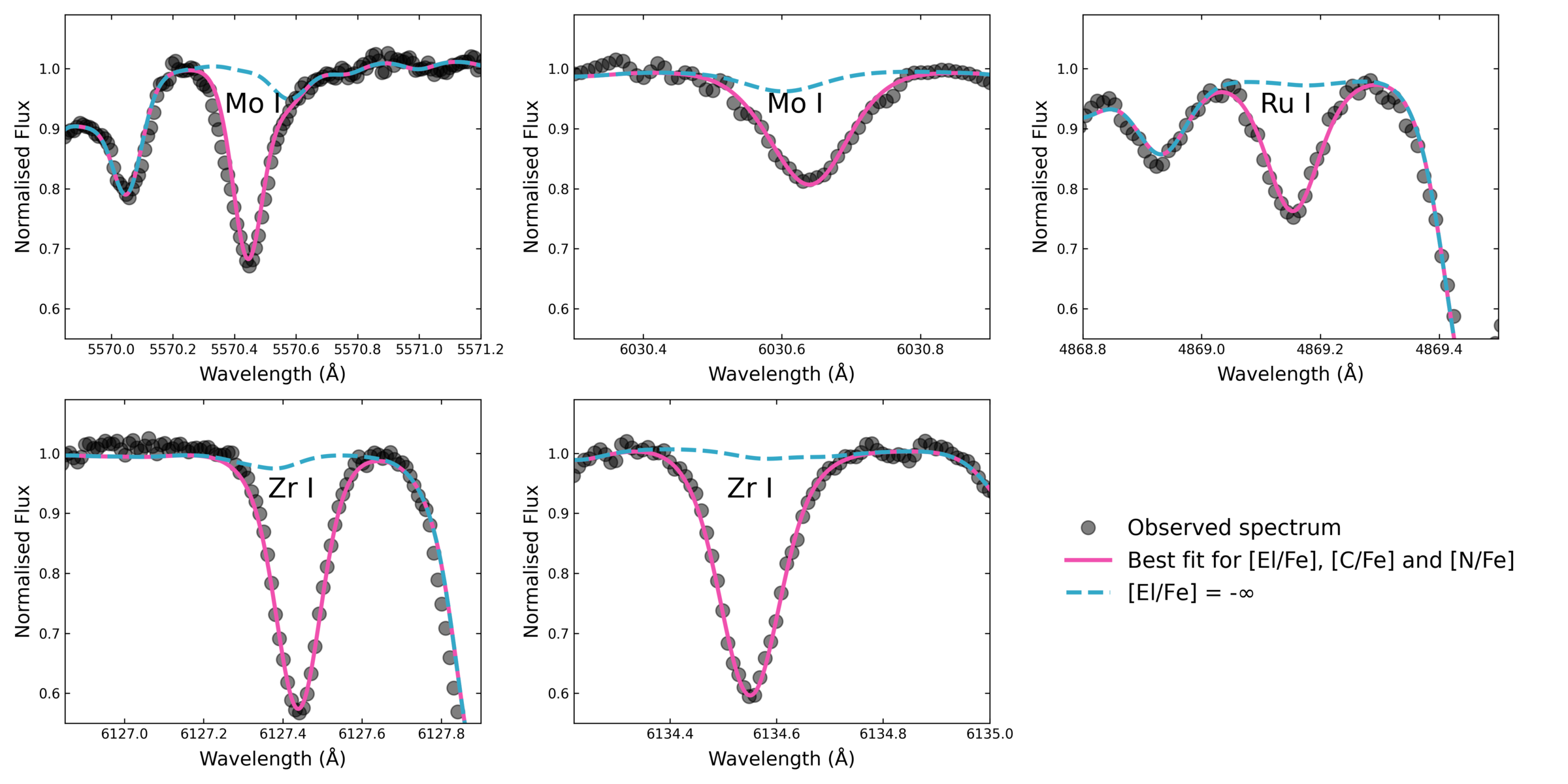}
\caption{Observed spectrum (black circles) for the star IC\,4756\,star\,\#2. In each panel, the element under consideration (Mo, Ru or Zr) is indicated. The magenta solid line represents our best fit for each studied element using the estimated [C/Fe] and [N/Fe]. The dashed teal line represents the synthetic spectrum computed without including that element.}
\label{fig:all_lines}
\end{figure*}

We adopted Mo abundances derived from the 5570 and 6030 \r{A} lines only when they were well-defined—specifically, when they appeared as relatively strong (EW $\gtrsim$ 7 m\AA), unsaturated, and sufficiently isolated features (see Fig.~\ref{fig:all_lines}). Atomic data for these transitions were sourced from the \textit{Gaia}-ESO line list \citep{Heiter2021_GESline}, incorporating updates from \citet{Magg2022} (see Table~\ref{table:atomicparams}). Following \citet{Mishenina2019}, these weak subordinate lines originate in deep atmospheric layers where LTE typically holds. Therefore, we performed our analysis under the LTE assumption, as departures are expected to be negligible and specific NLTE corrections are currently unavailable for these lines.

For the Ru determination, after testing several \ion{Ru}{i} lines (e.g., 4757.764 \r{A}, 4757.848 \r{A}, 4584.443 \r{A}, 5309.265 \r{A}, 5636.237 \r{A}), we adopted the \ion{Ru}{i} line at 4869.153 \r{A} (see Table~\ref{table:atomicparams} for the atomic parameters), classified as $Yes/Undecided$ by \citet{Heiter2021_GESline}. This line (see Fig.~\ref{fig:all_lines}) was also employed by \citet{Roriz2021} in their study of Ba stars. There is a small blend with the \ion{Si}{i} line at 4869.086 \r{A}, which we assume as negligible for our Ru determination, as we obtained a median sensitivity of -0.01 dex when setting the [Si/Fe] = 0.25 dex. Here, the sensitivity refers to the change in the derived abundances resulting from a variation in the assumed input stellar parameter or abundance. The $\log~gf$ and low-level excitation potential ($\chi^{\text{low}}_{\text{exc}}$) for this  \ion{Si}{i} line were determined in \citet{PehlivanRhodin2024}. 

Our Zr abundances were determined using the two \ion{Zr}{i} lines at 6127.440 \r{A} and 6134.550 \r{A} (see Fig.~\ref{fig:all_lines}), as previously done by \citet{Forsberg2019} and \citet{Mishenina2026}. These \ion{Zr}{i} lines are described by \citet{Velichko2010} as weak features with the presence of molecular blends. In general, \ion{Zr}{i} is more prone to being affected by NLTE effects than \ion{Zr}{ii}, and \citet{Velichko2010} showed that over-ionisation can produce abundance corrections of about +0.30 dex for strong \ion{Zr}{i} lines in RGB stars. In our analysis, we also explored several \ion{Zr}{ii} lines (e.g., 4050.220 \r{A},  4208.880 \r{A}, 4257.941 \r{A}, 4442.892 \r{A},  4496.862 \r{A},  5112.170 \r{A}), as well as the \ion{Zr}{i} lines at 4241.606 \r{A} and 4687.705 \r{A} used by \citet{Velichko2010}, and the \ion{Zr}{i} line at 6140.460 \r{A} used by \citet{Forsberg2019}. However, they failed to provide satisfactory fits for the observed spectra. Given that the \ion{Zr}{i} lines considered here are weak, as explained by \citet{Mishenina2026}, it remains uncertain whether NLTE corrections similar to those reported for stronger lines would apply in this case. The \ion{Zr}{i} line at 6127.440 \r{A} (\textit{Yes/Yes}) shows a small blend with the \ion{Fe}{i} line at 6127.3912 \r{A}. On the other hand, the line at 6134.550 \r{A} (\textit{Yes/Yes}) shows a small feature mainly due to CN molecules. The atomic data for these two lines is presented in Table~\ref{table:atomicparams}.

The elemental abundances were determined as the average of the values derived from individual spectral lines. As well as for the two \ion{Mo}{i} lines, the estimated [C/Fe] and [N/Fe] were adopted as input parameters for the spectral synthesis of the \ion{Ru}{i} and the \ion{Zr}{i}. The associated uncertainties were computed as $\sigma / \sqrt{n}$, where $\sigma$ represents the standard deviation among the measured lines and $n$ is the number of lines used. In cases where only a single line was available, we adopted a representative uncertainty of 0.04 dex, corresponding to the median uncertainty of the sample. For Ru, which was measured using only one line, the uncertainty was instead estimated from the dispersion of measurements within each OC. For clusters with only one star, we again adopted the median value of 0.04 dex. At the cluster level, abundances were computed as the mean of the stellar measurements within each OC, and the associated uncertainties were estimated as the standard deviation of the cluster divided by $\sqrt{N}$, where $N$ is the number of stars with available measurements. For clusters with only one star, we adopted median uncertainties of 0.05 dex for [Mo/Fe] and [Zr/Fe], and 0.04 dex for [Ru/Fe]. The sensitivities to stellar parameters and C and N abundances are reported in Table~\ref{table:sensitivities}.

\section{Results and Discussion}\label{sec:results}

In this section, we present the elemental abundances of Mo, Ru, and Zr derived in this work, compare our estimates with previous studies, and discuss the associated uncertainties. 
The derived abundances for Mo, Ru and Zr for individual stars are presented in  Table~\ref{table:Abund_stars}, while the corresponding mean abundances for each OC are listed in Table~\ref{table:Abund_OCs}. In total, we obtained Mo abundances for 80 stars across 30 OCs, Ru abundances for 66 stars in 26 OCs, and Zr abundances for 81 stars in 30 OCs. All results were visually inspected to ensure the reliability and quality of the fit.

\begin{table*}[ht!]
\caption{Mean [Mo/Fe], [Ru/Fe] and [Zr/Fe] obtained for each OC as well as the number of stars considered for the estimation of each element.}     
\label{table:Abund_OCs}    
\centering                        
\begin{tabular}{lcccccc}
\hline\hline
OC & [Mo/Fe] & $N_{\mathrm{stars}}^{\mathrm{Mo}}$ & [Ru/Fe] & $N_{\mathrm{stars}}^{\mathrm{Ru}}$  & [Zr/Fe] & $N_{\mathrm{stars}}^{\mathrm{Zr}}$  \\
 & (dex) &  & (dex) &  & (dex) &  \\
\hline
Alessi 1 & $0.08 \pm 0.02$ & 4 & $0.10 \pm 0.03$ & 4 & $-0.06 \pm 0.01$ & 4 \\
Alessi 161 & $0.06 \pm 0.05$ & 1 & $0.13 \pm 0.04$ & 1 & $-0.04 \pm 0.05$ & 1 \\
Alessi-Teutsch 11 & $0.02 \pm 0.05$ & 1 & $0.05 \pm 0.04$ & 1 & $-0.13 \pm 0.05$ & 1 \\
Basel 11b & $0.06 \pm 0.03$ & 3 & $-$ & 0 & $0.01 \pm 0.04$ & 3 \\
COIN-Gaia 30 & $0.05 \pm 0.05$ & 1 & $0.02 \pm 0.04$ & 1 & $0.00 \pm 0.05$ & 1 \\
Collinder 350 & $0.12 \pm 0.05$ & 1 & $-$ & 0 & $0.07 \pm 0.05$ & 1 \\
Collinder 463 & $0.06 \pm 0.02$ & 2 & $0.14 \pm 0.02$ & 2 & $-0.09 \pm 0.03$ & 2 \\
Gulliver 51 & $-0.04 \pm 0.05$ & 1 & $0.02 \pm 0.04$ & 1 & $-0.14 \pm 0.05$ & 1 \\
IC 4756 & $0.11 \pm 0.02$ & 12 & $0.13 \pm 0.02$ & 11 & $0.00 \pm 0.02$ & 12 \\
LP 1800 & $0.04 \pm 0.02$ & 3 & $0.05 \pm 0.01$ & 2 & $-0.06 \pm 0.02$ & 3 \\
NGC 2437 & $0.09 \pm 0.03$ & 5 & $0.13 \pm 0.01$ & 4 & $-0.05 \pm 0.04$ & 5 \\
NGC 2509 & $-0.09 \pm 0.05$ & 1 & $-$ & 0 & $-0.14 \pm 0.05$ & 1 \\
NGC 2548 & $0.07 \pm 0.06$ & 3 & $0.11 \pm 0.10$ & 2 & $-0.05 \pm 0.06$ & 3 \\
NGC 2632 & $-0.03 \pm 0.02$ & 2 & $0.07 \pm 0.02$ & 2 & $-0.18 \pm 0.04$ & 2 \\
NGC 2682 & $-0.01 \pm 0.02$ & 2 & $0.07 \pm 0.01$ & 2 & $-0.15 \pm 0.01$ & 2 \\
NGC 6800 & $0.16 \pm 0.05$ & 1 & $0.22 \pm 0.04$ & 1 & $0.11 \pm 0.05$ & 1 \\
NGC 6991 & $0.11 \pm 0.03$ & 5 & $0.14 \pm 0.02$ & 5 & $-0.01 \pm 0.02$ & 5 \\
NGC 7086 & $0.02 \pm 0.01$ & 2 & $0.07 \pm 0.01$ & 2 & $-0.10 \pm 0.01$ & 2 \\
NGC 7209 & $-0.01 \pm 0.05$ & 1 & $0.03 \pm 0.04$ & 1 & $-0.11 \pm 0.05$ & 1 \\
NGC 752 & $0.01 \pm 0.02$ & 3 & $0.13 \pm 0.03$ & 3 & $-0.04 \pm 0.01$ & 3 \\
Ruprecht 171 & $-0.09 \pm 0.01$ & 4 & $-0.06 \pm 0.04$ & 4 & $-0.28 \pm 0.02$ & 4 \\
Stock 2 & $0.10 \pm 0.03$ & 8 & $0.13 \pm 0.02$ & 6 & $-0.01 \pm 0.03$ & 8 \\
Tombaugh 5 & $0.11 \pm 0.08$ & 3 & $0.02 \pm 0.04$ & 1 & $-0.03 \pm 0.06$ & 3 \\
UBC 141 & $0.00 \pm 0.05$ & 1 & $0.04 \pm 0.04$ & 1 & $-0.15 \pm 0.05$ & 1 \\
UBC 169 & $0.08 \pm 0.05$ & 1 & $-$ & 0 & $0.07 \pm 0.05$ & 1 \\
UBC 194 & $0.02 \pm 0.05$ & 1 & $-$ & 0 & $0.02 \pm 0.05$ & 1 \\
UBC 577 & $0.13 \pm 0.03$ & 3 & $0.06 \pm 0.04$ & 3 & $0.07 \pm 0.06$ & 4 \\
UBC 60 & $0.06 \pm 0.05$ & 1 & $0.15 \pm 0.04$ & 1 & $-0.04 \pm 0.05$ & 1 \\
UPK 219 & $0.11 \pm 0.05$ & 1 & $0.11 \pm 0.04$ & 1 & $0.03 \pm 0.05$ & 1 \\
UPK 84 & $0.14 \pm 0.03$ & 3 & $0.21 \pm 0.02$ & 3 & $0.05 \pm 0.01$ & 3 \\
\hline
\end{tabular}
\end{table*}

\begin{figure}
   \centering
   \includegraphics[width=0.85\hsize]{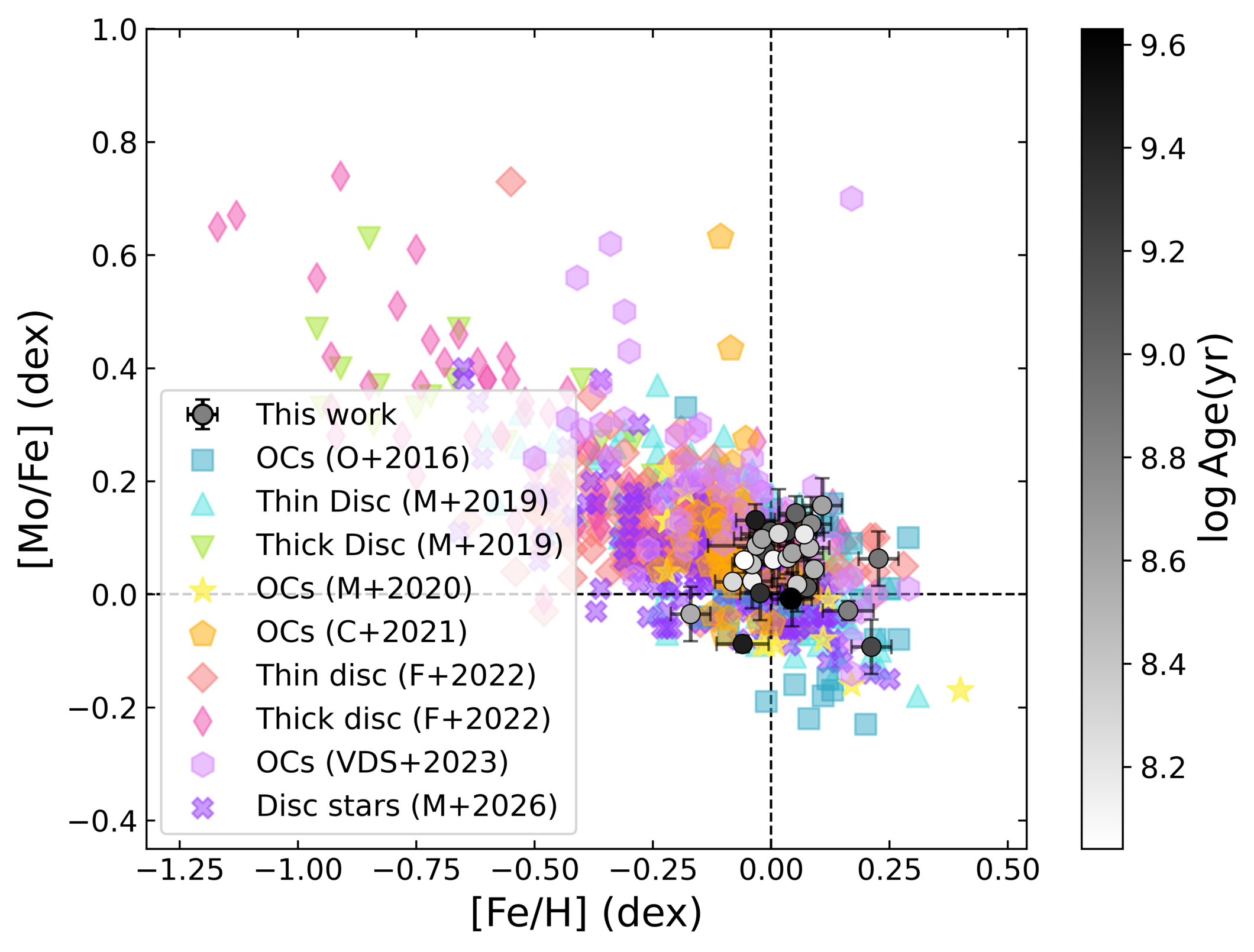}
   \includegraphics[width=0.85\hsize]{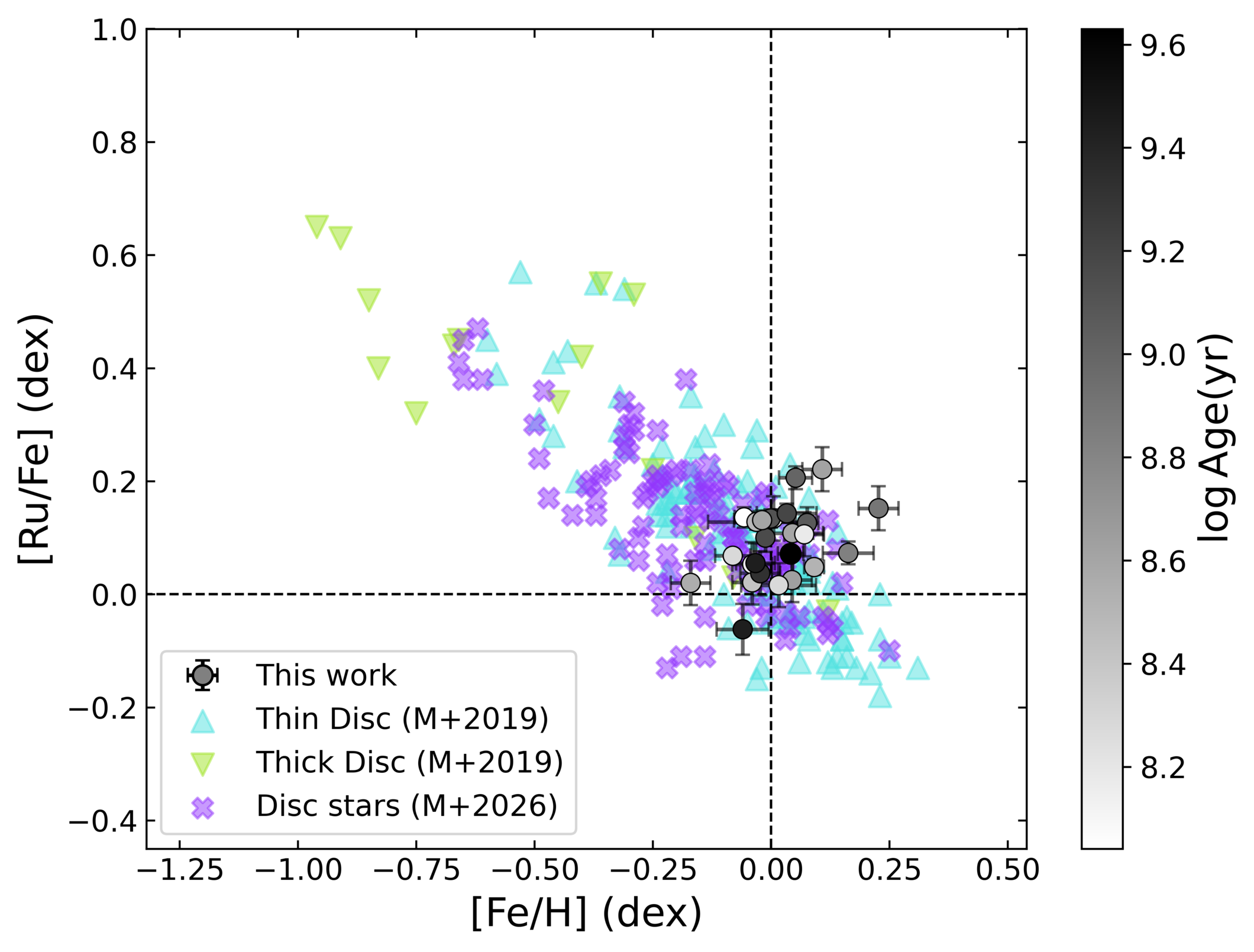}
   \includegraphics[width=0.85\hsize]{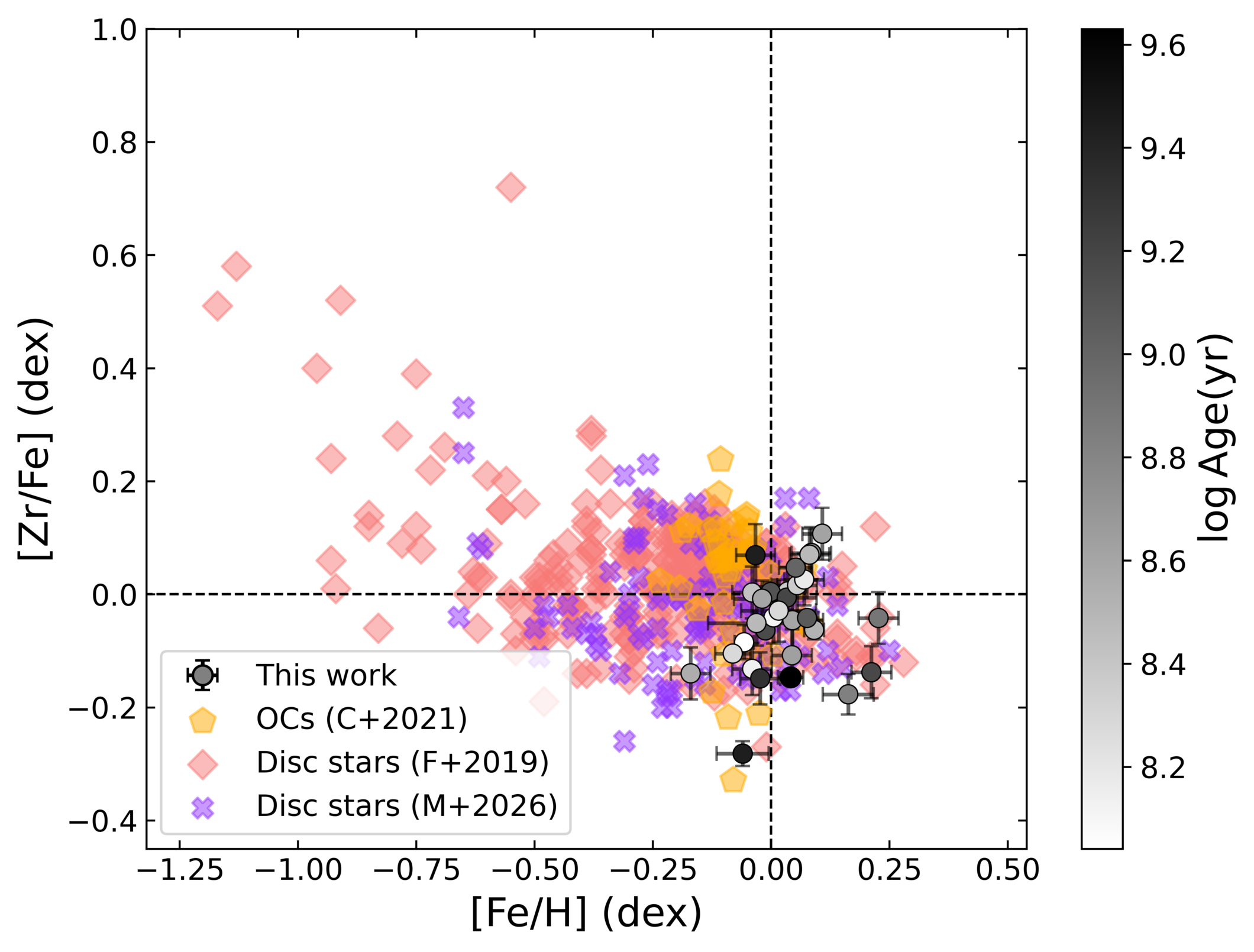}
      \caption{[Mo/Fe] (upper panel), [Ru/Fe] (middle panel) and [Zr/Fe] (bottom panel) with respect to [Fe/H]. For each OC, our results are presented as circles, colour-coded by its age. The teal squares represent the OCs from \citet[][O+2016]{Overbeek2016}, the upward cyan triangle and the downward green triangle represent the thin and thick disc stars from \citet[][M+2019]{Mishenina2019}. The yellow stars are the OCs from \citet[][M+2020]{Mishenina2020}, while the orange pentagons are the OCs from \citet[][C+2021]{Casamiquela2021}. The red rotated squares and the pink diamonds are the thin and thick disc stars from \citet[][F+2022]{Forsberg2022} respectively in the case of [Mo/Fe], while for [Zr/Fe], the red rotated squares represent all the disc stars from \citet[][F+2019]{Forsberg2019}. The light purple hexagons are the OCs from \citet[][VDS+2023]{VanderSwaelmen2023} and the dark purple crosses are the thin disc giants from \citet[][M+2026]{Mishenina2026}. The dashed black lines that go through [0, 0] indicate the solar value.}
         \label{fig:ElFe_FeH}
\end{figure}

In Fig.~\ref{fig:ElFe_FeH} we present our results for  [Mo/Fe], [Ru/Fe] and [Zr/Fe] with respect to [Fe/H]. In this figure, we also present the results for thin and thick disc dwarfs \citep{Mishenina2019}, thin and thick disc giants \citep{Forsberg2019, Forsberg2022} and thin disc giants \citep[][excluding HD~6833 due to its radial velocity (RV) measurement\footnote{HD~6833 has a reported RV of -243.38 kms$^{-1}$ \citet{Gaia_2023Vallenari} which does not seem to be consistent with thin disc kinematics \citep[see][]{Anguiano2020}.}]{Mishenina2026}. Additionally, we included the OCs from \citet{Overbeek2016, Mishenina2020, Casamiquela2021, VanderSwaelmen2023}, excluding the OCs King~1 and NGC~2266 from \citet{Casamiquela2021}, as their reported [Mo/H] abundances are approximately 0.5 dex, corresponding to deviations of about 10$\sigma$ from the mean distribution, and are associated with comparatively large uncertainties.

\begin{figure}
   \centering
   \includegraphics[width=\hsize]{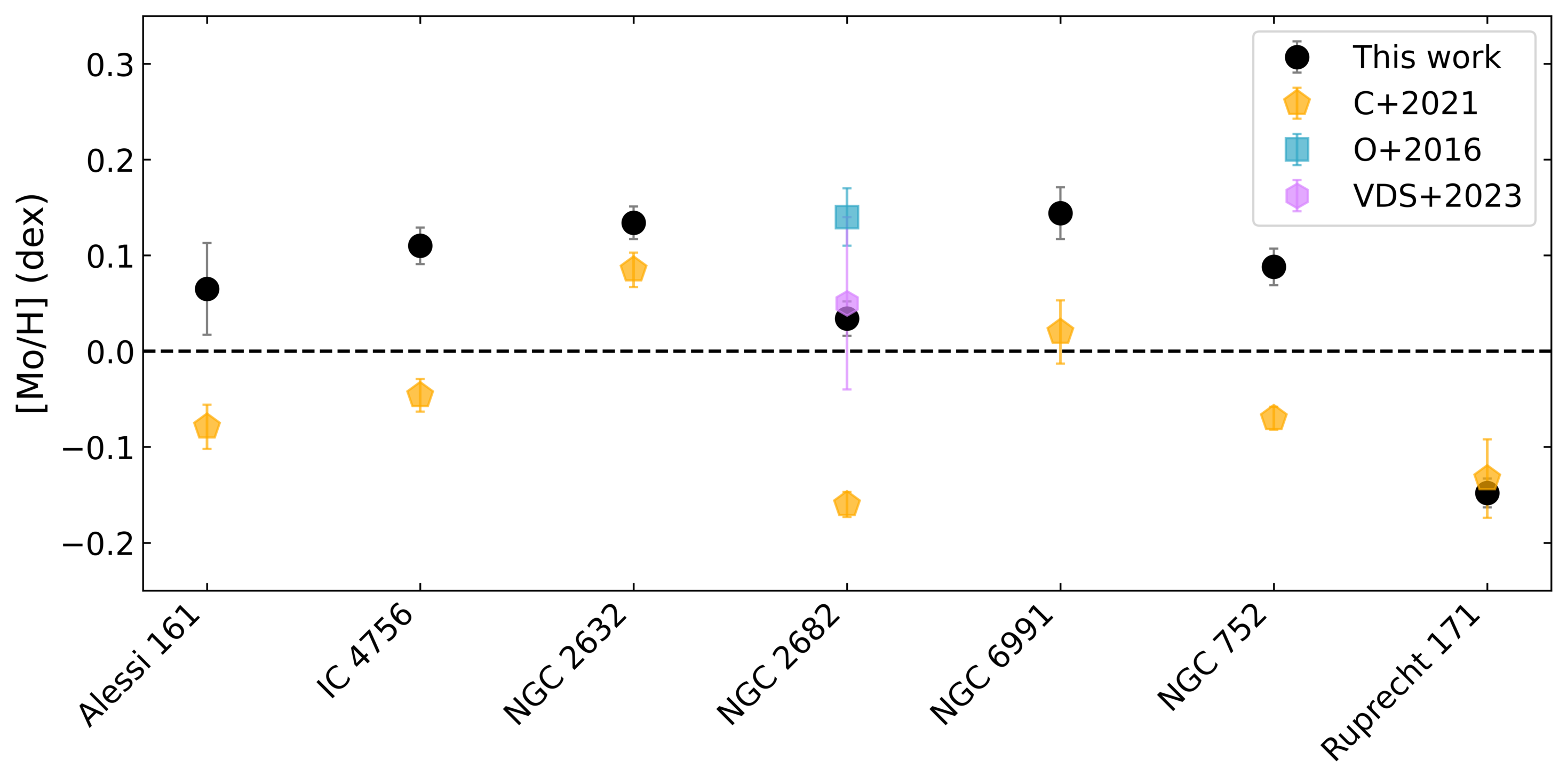}
   \includegraphics[width=\hsize]{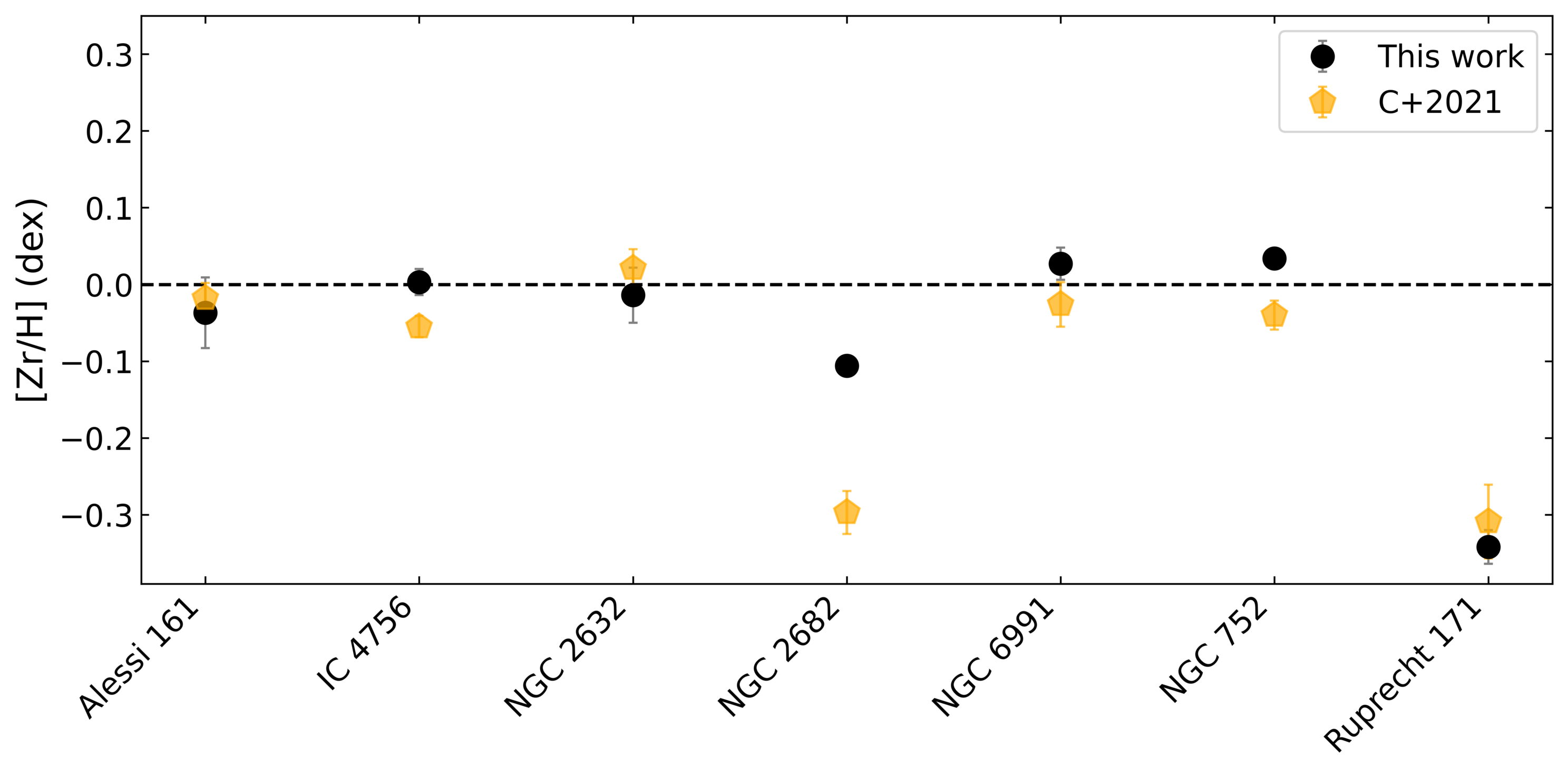}
   \includegraphics[width=\hsize]{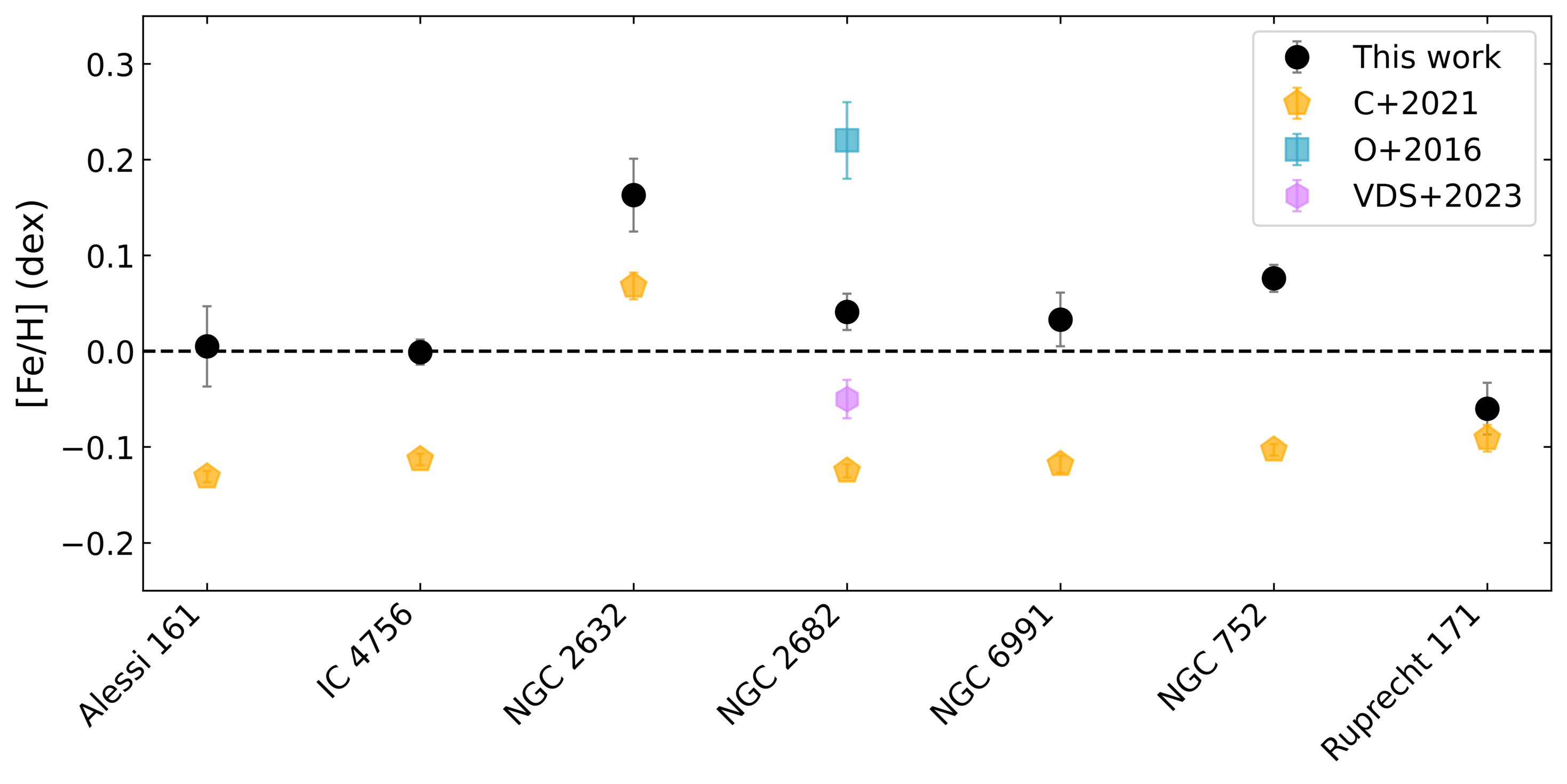}
      \caption{Comparison of the values of [Mo/H] (upper panel), [Zr/H] (middle panel), and [Fe/H] (bottom panel) derived in this study (black circles) for the OCs in common with \citet[][C+2021]{Casamiquela2021}, \citet[][O+2016]{Overbeek2016}, and \citet[][VDS+2023]{VanderSwaelmen2023}, represented by orange pentagons, teal squares, and light purple hexagons, respectively, with their corresponding errors. The [Fe/H] values labelled as ``This work" are adopted from \citetalias{SPA_DalPonte2025}.}
         \label{fig:compare_OCs}
\end{figure}

The values from the literature were renormalised to our adopted solar scale (see Sect.~\ref{sec:specanalyses}). For the three planes in Fig.~\ref{fig:ElFe_FeH}, there is an overall agreement between our results and those from the literature for the restricted metallicity range covered by our sample. In all three panels, the old OC Ruprecht~171 with [Fe/H]~$\sim$~-0.06 dex exhibits abundances lower than expected when compared with OCs of similar [Fe/H]. However, when compared with \citet{Casamiquela2021} (see Fig.~\ref{fig:compare_OCs}) they obtained [Mo/Fe]~=~-0.04~$\pm$~0.04~dex and [Zr/Fe]~=~-0.22~$\pm$~0.05~dex which are consistent within the errors with our values of [Mo/Fe]~=~-0.09~$\pm$~0.01~dex and [Zr/Fe]~=~-0.28~$\pm$~0.02~dex. 

Since the abundances from \citet{Casamiquela2021} were derived from different spectra using independent methodology and linelist, this agreement supports the robustness of the low abundances, in particular the [Zr/Fe], and makes it unlikely they are caused solely by observational limitations or systematic effects specific to our analysis. We estimated the guiding radius following \citet{Spina2021}, using the \texttt{galpy} package and assuming the MW2014 potential for the Milky Way \citep{Bovy2015}. We also examined the orbital parameters reported by \citet{Tarricq2021} for  Ruprecht~171, but found no clear kinematical feature that could readily account for its peculiar chemical pattern. A more detailed kinematical analysis would be required to explore this possibility further, but it lies beyond the scope of this work.

Also, in all three panels, the intermediate-age OC NGC~6800 with [Fe/H]~$\sim$~0.11~dex yields the highest abundances. For this cluster we obtained [Mo/Fe]~=~0.16~$\pm$~0.05~dex, [Ru/Fe]~=~0.22~$\pm$~0.04~dex and [Zr/Fe]~=~0.11~$\pm$~0.05~dex. The three more metal-rich OCs (UBC~60 with [Fe/H]~$\sim$~0.23~dex, NGC~2509 with [Fe/H]~$\sim$~0.21~dex and NGC~2632 with [Fe/H]~$\sim$~0.16~dex) seem to lie slightly off the distribution with the case of the [Ru/Fe]~=~0.15~$\pm$~0.04~dex for UBC~60 being the most noticeable. When comparing our results with \citet{Casamiquela2021} for NGC~2632, they found [Mo/Fe]~=~0.02~$\pm$~0.02~dex and [Zr/Fe]~=~-0.05~$\pm$~0.03~dex with respect to our 
[Mo/Fe]~=~-0.03~$\pm$~0.02~dex and [Zr/Fe]~=~-0.18~$\pm$~0.04~dex. [Mo/Fe] is consistent within 2$\sigma$ while [Zr/Fe] exhibits a deviation of about 3$\sigma$.

We compare in Fig.~\ref{fig:compare_OCs} the results for our OCs in common with \citet{Casamiquela2021}, including NGC~2682 (M67) that is also in common with \citet{Overbeek2016} and \citet{VanderSwaelmen2023}. We present the different estimates for [Mo/H], [Zr/H] and [Fe/H]\footnote{[El/Fe] = [El/H] - [Fe/H]}. When comparing with \citet{Casamiquela2021}, we found our estimates to be systematically higher for [Mo/H] and [Fe/H], with the exception of Ruprecht~171, which is in good agreement with their reported value. Regarding [Zr/H], we find no significant systematic offset between our results and those of \citet{Casamiquela2021}. Specifically, the clusters Alessi~161, NGC~2632, NGC~6991, and Ruprecht~171 show good agreement within the uncertainties, with discrepancies generally remaining below 0.1 dex, except for NGC~2682. 
Notice that our [Mo/H] and [Fe/H] estimates for NGC~2682 are approximately 0.1 dex and 0.2 dex lower, respectively, than those reported by \citet{Overbeek2016}. In contrast, our [Mo/H] value for this cluster is in good agreement with \citet{VanderSwaelmen2023} within the uncertainties, though a difference of $\sim$0.1 dex persists in [Fe/H].

To investigate the nucleosynthetic origin of Mo and Ru, we compare their abundances with those of elements whose nucleosynthesis production channels are well established \citep{Hansen2014, Mishenina2019}. For this purpose, we analyse the slopes defined by the absolute abundances A(El)\footnote{[El/H] = A(El) - A(El)$_\odot$ = $\log \left(\frac{n_{\text{El}}}{n_\text{H}}\right)_*-\log \left(\frac{n_{\text{El}}}{n_\text{H}}\right)_\odot$} between the elements measured in this work (Mo, Ru, and Zr) and those reported by \citetalias{SPA_DalPonte2025}, in particular, the neutron-capture elements Sr, Y (predominantly produced by the s-process) and Eu (mainly produced by the r-process). 

Since the comparison is performed using absolute abundances, it is not affected by the systematic offsets introduced by differences in the adopted solar abundance scale or by the use of Fe as reference element in abundance ratios.

For obtaining the slopes, we follow the procedure described by \citet{Nunnari2025}. Using a Bayesian approach to infer model parameters (slope $m$ and intercept $b$ following the equation $y = mx + b$), along with their uncertainties. For a dataset $D$ and parameters $\theta$, the posterior probability is given by:
$$
p(\theta \mid D) \propto \mathcal{L}(D \mid \theta) \pi(\theta)
$$

Where $\pi(\theta)$ represents the prior distribution and $\mathcal{L}$ is the likelihood. In practice, we used the log-posterior:
$$
\log p(\theta \mid D)=\log \mathcal{L}(D \mid \theta)+\log \pi(\theta)+\text { const } 
$$

Posterior sampling is performed using the affine-invariant Markov Chain Monte Carlo (MCMC) algorithm from the \texttt{emcee}\footnote{https://github.com/dfm/emcee} library  \citep{emcee}, which allows us to obtain maximum a posteriori (MAP) estimates and credible intervals corresponding to the 68$\%$ region of the marginal distributions.

We specifically use mixture models \citep{ForemanMackey2014}, where the likelihood of each data point is expressed as a weighted combination of a foreground (signal) and a background (outlier) component. Let $Q$ denote the probability that a data point arises from the background, called the mixing fraction. Then, for a datum $(x_i, y_i)$ with uncertainties $(\sigma_{x,i}, \sigma_{y,i})$, the individual likelihood is:
$$
\mathcal{L}_i=(1-Q) \mathcal{L}_{f g}\left(x_i, y_i, \sigma_{x, i}, \sigma_{y, i} \mid \theta_{f g}\right)+Q \mathcal{L}_{b g}\left(x_i, y_i, \sigma_{x, i}, \sigma_{y, i} \mid \theta_{b g}\right)
$$

And, the total log-likelihood is:

$$
\log \mathcal{L}=\sum_i \log \mathcal{L}_i .
$$

Hence, the mixing fraction $Q$ controls the influence of the background model, making the inference less sensitive to outliers. In Gaussian implementations, each component is typically modelled as a normal distribution with its own mean and variance, and may also include an additional ``jitter'' term, added in quadrature, to account for potential underestimation of errors. For numerical stability, the evaluation of $\log \mathcal{L}$ should employ log-sum-exp techniques, which prevent underflow when one component contributes much less than the other.

Convergence of the MCMC chains was assessed using estimates of the autocorrelation time and by visual inspection of the trace plots. To quantify the quality of the fits, we report the mean squared error and the Akaike Information Criterion (AIC). The AIC provides a measure for model comparison that balances goodness-of-fit against model complexity. It is defined as:
$$
\mathrm{AIC} = 2k - 2 \log \mathcal{L}_\mathrm{max}
$$

Where $k$ is the number of model parameters and $\mathcal{L}_\mathrm{max}$ is the maximised likelihood. Models with lower AIC values are generally preferred, as they indicate a better compromise between fit quality and model simplicity.

The use of the mixture models instead of a conventional linear least-squares fitting was primarily motivated by the fact that the latter is only appropriate when the uncertainties are negligible along one axis and can be described by Gaussian distributions of known variance along the other \citep{Hogg2010}. These requirements do not hold for our observational data. Additionally, as it is also explained by \citet{Hogg2010}, the mixture model provides a more appropriate treatment of outliers. Manually excluding these points or applying iterative sigma clipping would require subjective choices concerning the rejection threshold and the stage at which the rejection is performed. In contrast, the mixture model provides a probabilistic treatment of the full dataset.

\begin{figure*}
   \centering
\includegraphics[width=\hsize]{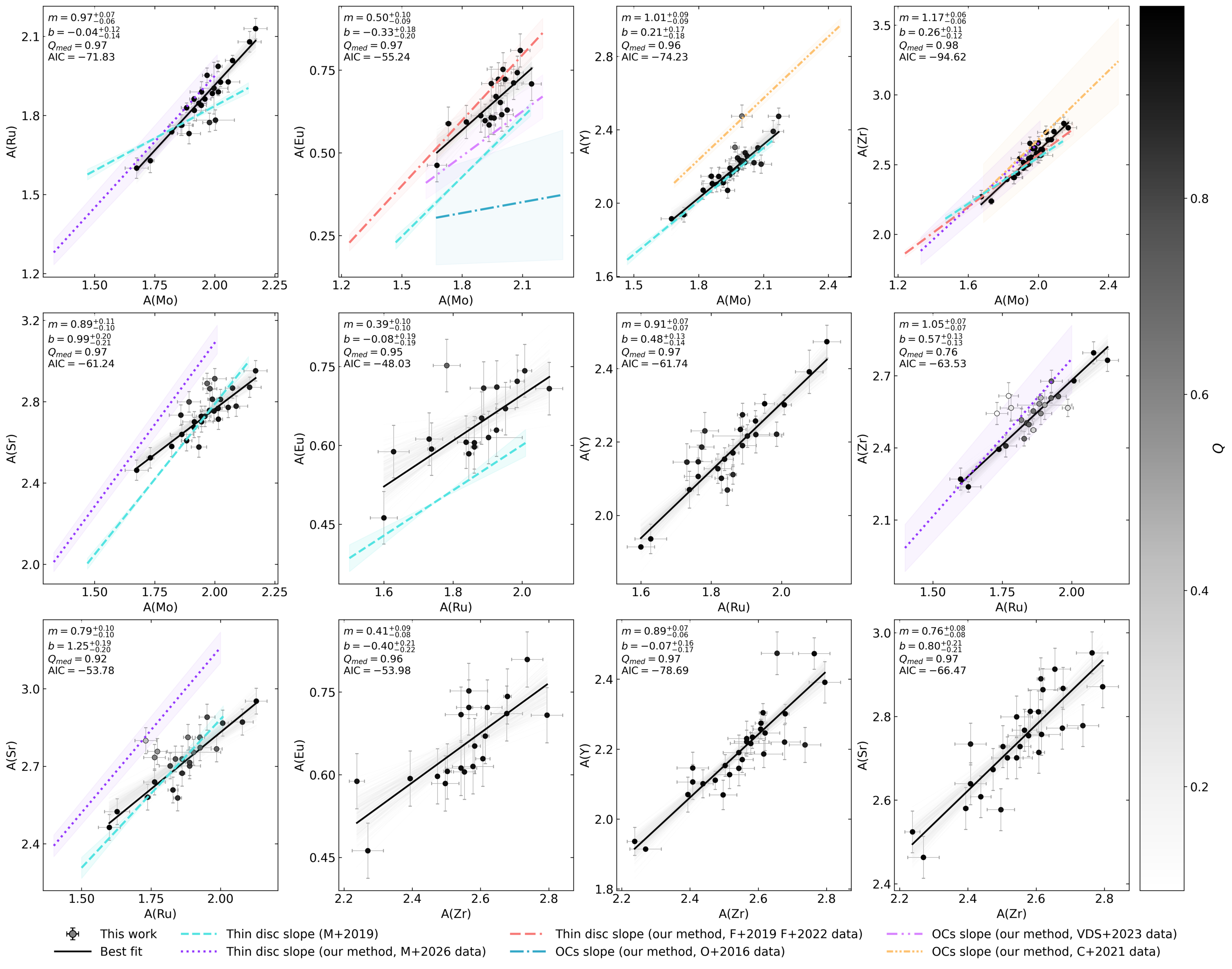}
\caption{Linear trends between the absolute abundances  A(El$_{\text{y-axis}}$) vs A(El$_{\text{x-axis}}$)  of the measured elements. The OCs of this work are shown as circles, colour-coded in grayscale according to their quality factor $Q$. The black solid line represents the best-fit relation derived using our method described in Sect.~\ref{sec:results}. For each panel, the fitted slope 
$m$ and intercept $b$, along with their uncertainties, as well as the median quality factor $Q_{\text{med}}$ and the AIC, are reported. For comparison, the slopes reported in \citet{Mishenina2019} are plotted as cyan dashed lines. Additionally, we fitted using our method the data from \citet{Mishenina2026} shown as a dotted dark purple line and the thin disc stars from \citet{Forsberg2019, Forsberg2022} as red dashdashdotted lines. The OCs from \citet{Overbeek2016}, \citet{VanderSwaelmen2023} and \citet{Casamiquela2016} as teal dash-dotted lines, light purple dashdotdotted lines and yellow densely dashdotdotted lines, respectively. The best-fit parameters for this work and for the literature are reported in  Table~\ref{table:fits}.}
\label{fig:slopes}
\end{figure*}

Then, using this approach, we derive the slopes for our OCs as shown in Fig.~\ref{fig:slopes}, and compare them with the slopes reported by \citet{Mishenina2019}. Since no slopes were provided for the thin disc stars in \citet{Forsberg2019,Forsberg2022} and \citet{Mishenina2026}, nor for the OCs in \citet{Overbeek2016},  \citet{Casamiquela2021} and \citet{VanderSwaelmen2023}, we applied the same method to these samples. The resulting slopes for the available elements are reported in Table~\ref{table:fits}.

When comparing A(Ru) and  A(Mo), we found a strong correlation between Mo and Ru, consistent with the fact that both elements have isotopes produced through a combination of p-, r-, and s-process nucleosynthesis. Our results are in good agreement with \citet{Mishenina2026} for thin disc giants, while showing a discrepancy with the slope for the thin disc dwarfs reported in \citet{Mishenina2019}. Offsets between dwarfs and giants have been previously reported and may arise as a consequence of the analysis itself \citep[see][for a discussion on this issue]{Dorazi2009}.

For the slopes between A(Eu) and A(Mo), we find a slope of $\sim$0.5, indicating that the r-process alone is insufficient to account for the observed Mo abundances, as expected from the isotopic contributions summarised in Table~\ref{table:mo_ru_isotopes_full}. Our results are consistent with those reported for thin disc stars by \citet{Mishenina2019} and  \citet{Forsberg2019,Forsberg2022}, as well as for the OCs by \citet{VanderSwaelmen2023}. However, while the A(Mo) range derived by \citet{Overbeek2016} agrees with ours, we find a discrepancy in A(Eu). Investigating the origin of this difference is beyond the scope of this work, but it may arise from differences in the Eu abundance determination in \citet{Overbeek2016}. Consistently, we derived a slope for A(Eu) versus A(Ru) of $\sim$ 0.39, also in agreement with those of \citet{Mishenina2019}, while for A(Eu) versus A(Zr) we obtain $\sim$ 0.42. For the s-process elements (Sr, Y, Zr) we find slopes near unity when compared with Mo and Ru and among themselves, implying similar enrichment timescales. These results agree with findings for thin-disc stars by \citet{Mishenina2019,Mishenina2026} and \citet{Forsberg2019,Forsberg2022}, as well as with the OCs reported by \citet{Casamiquela2021}, i.e., OCs show slopes consistent with field stars. The offsets in fitted parameters between studies likely arise from differences in spectroscopic analysis, e.g., line selection, atomic data, and analysis code \citep[see][for a more complete discussion on this topic]{Jofre2017, Jofre2019}. Offsets relative to \citet{Mishenina2019,Mishenina2026} and \citet{Forsberg2019,Forsberg2022} may additionally reflect that OCs are, on average, younger than field stars at the same metallicity.

\begin{table*}
\centering
\caption{Linear fit parameters for the equation $y = mx + b$ using the fitting procedure described in Sect.~\ref{sec:results}. The error is reported as the average of the upper and lower errors. }
\label{table:fits}
\begin{tabular}{llcc}
\hline\hline
Source & Relation & $m$ & $b$ \\
 & A(El$_{\text{y-axis}}$) vs. A(El$_{\text{x-axis}}$) &  & (dex) \\
\hline
\multirow{12}{*}{This work} 
  & A(Ru) vs. A(Mo) & 0.97 $\pm$ 0.07 & -0.04 $\pm$ 0.13 \\
  & A(Eu) vs. A(Mo) & 0.50 $\pm$ 0.10 & -0.33 $\pm$ 0.19 \\
  & A(Eu) vs. A(Ru) & 0.39 $\pm$ 0.10 & -0.08 $\pm$ 0.19 \\
  & A(Eu) vs. A(Zr) & 0.41 $\pm$ 0.09 & -0.40 $\pm$ 0.22 \\
  & A(Sr) vs. A(Mo) & 0.89 $\pm$ 0.11 & 0.99 $\pm$ 0.21 \\
  & A(Sr) vs. A(Ru) & 0.79 $\pm$ 0.10 & 1.25 $\pm$ 0.20 \\
  & A(Sr) vs. A(Zr) & 0.76 $\pm$ 0.08 & 0.80 $\pm$ 0.21 \\
  & A(Y) vs. A(Mo) & 1.01 $\pm$ 0.09 & 0.21 $\pm$ 0.18 \\
  & A(Y) vs. A(Ru) & 0.91 $\pm$ 0.07 &  0.48 $\pm$ 0.14 \\
  & A(Y) vs. A(Zr) & 0.89 $\pm$ 0.07 & -0.07 $\pm$ 0.17 \\
  & A(Zr) vs. A(Mo) & 1.17 $\pm$ 0.06  & 0.26 $\pm$ 0.12 \\
  & A(Zr) vs. A(Ru) & 1.05 $\pm$ 0.07  & 0.57 $\pm$ 0.13 \\

\hline

\citet{Overbeek2016} & A(Eu) vs. A(Mo) & 0.11 $\pm$ 0.17 & 0.12 $\pm$ 0.35 \\
\hline
\multirow{2}{*}{\citet{Casamiquela2021}} 
    & A(Y) vs. A(Mo)  &  1.12 $\pm$ 0.03 & 0.22 $\pm$ 0.06\\
    & A(Zr) vs. A(Mo) & 1.23 $\pm$ 0.25 & 0.23 $\pm$ 0.51 \\
\hline
\multirow{2}{*}{\citet{Forsberg2019,Forsberg2022}}     & A(Eu) vs. A(Mo)  & 0.66 $\pm$ 0.04 & -0.59 $\pm$ 0.07 \\
    & A(Zr) vs. A(Mo) & 0.93 $\pm$ 0.04 & 0.71 $\pm$ 0.07 \\
\hline
\citet{VanderSwaelmen2023} & A(Eu) vs. A(Mo)  & 0.45 $\pm$  0.06 & -0.32  $\pm$  0.12\\
\hline
\multirow{2}{*}{\citet{Mishenina2026}} 
    & A(Ru) vs. A(Mo)  & 1.00 $\pm$ 0.07 & -0.05 $\pm$ 0.13 \\
    & A(Sr) vs. A(Mo)  & 1.61 $\pm$ 0.07 & -0.13 $\pm$ 0.12 \\
    & A(Sr) vs. A(Ru)  & 1.28 $\pm$ 0.06 & 0.60 $\pm$ 0.11\\
    & A(Zr) vs. A(Mo) & 1.15 $\pm$ 0.15 & 0.35 $\pm$ 0.12 \\
    & A(Zr) vs. A(Ru) & 1.31 $\pm$ 0.14 & 0.15 $\pm$ 0.24 \\

\hline
\end{tabular}
\end{table*}

\begin{figure}
   \centering
   \includegraphics[width=0.78\hsize]{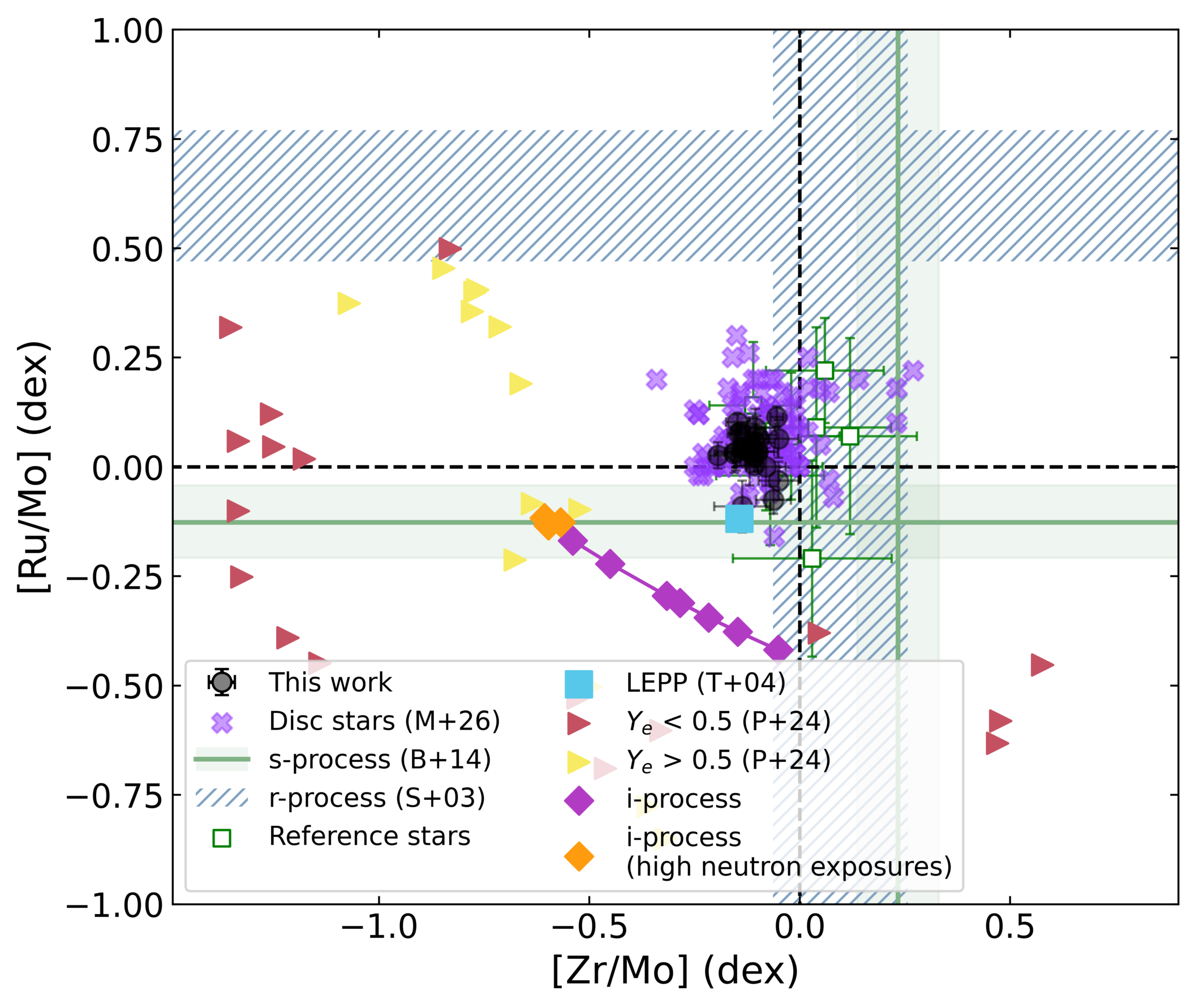}
   \includegraphics[width=0.78\hsize]{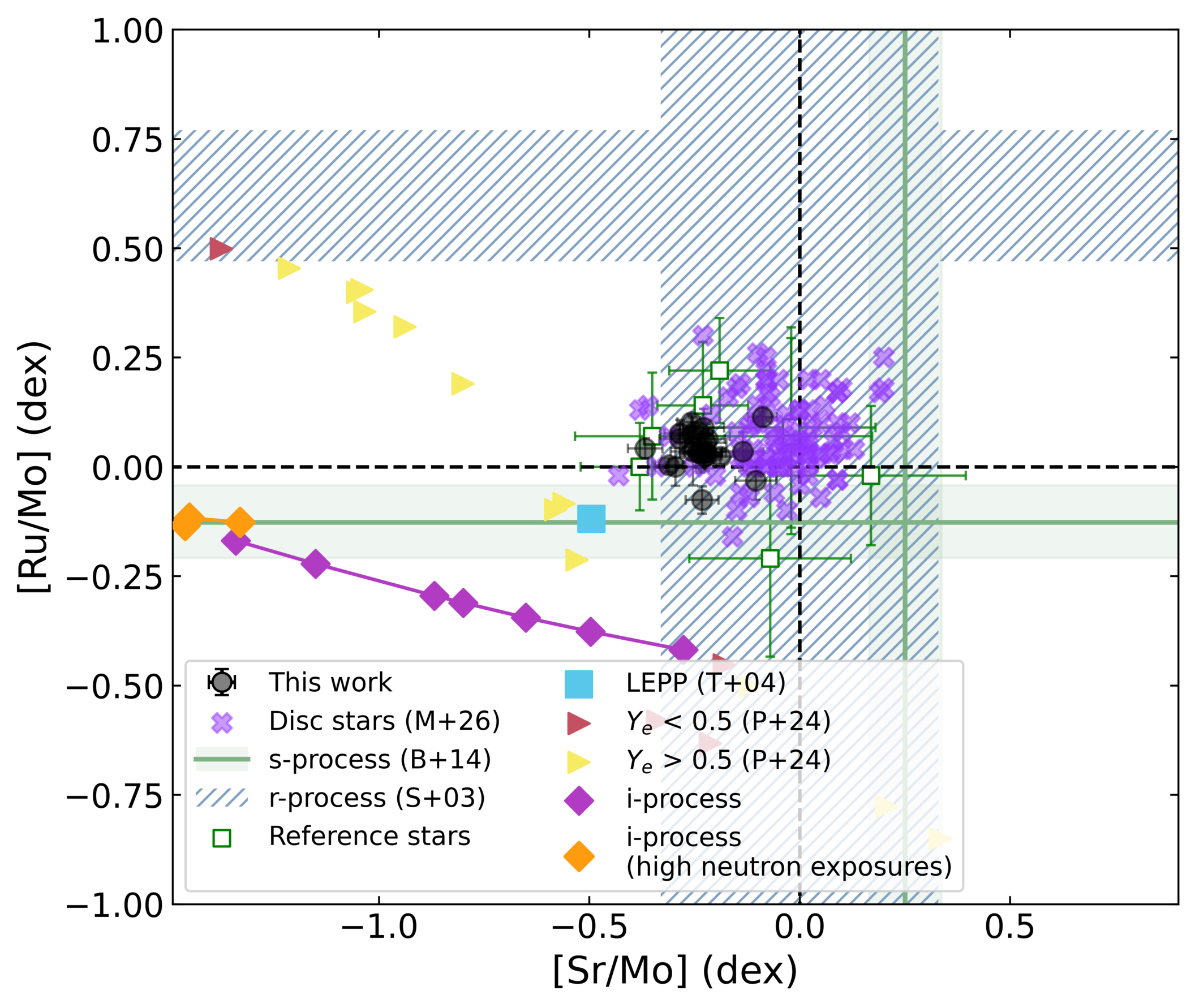}
   \includegraphics[width=0.78\hsize] {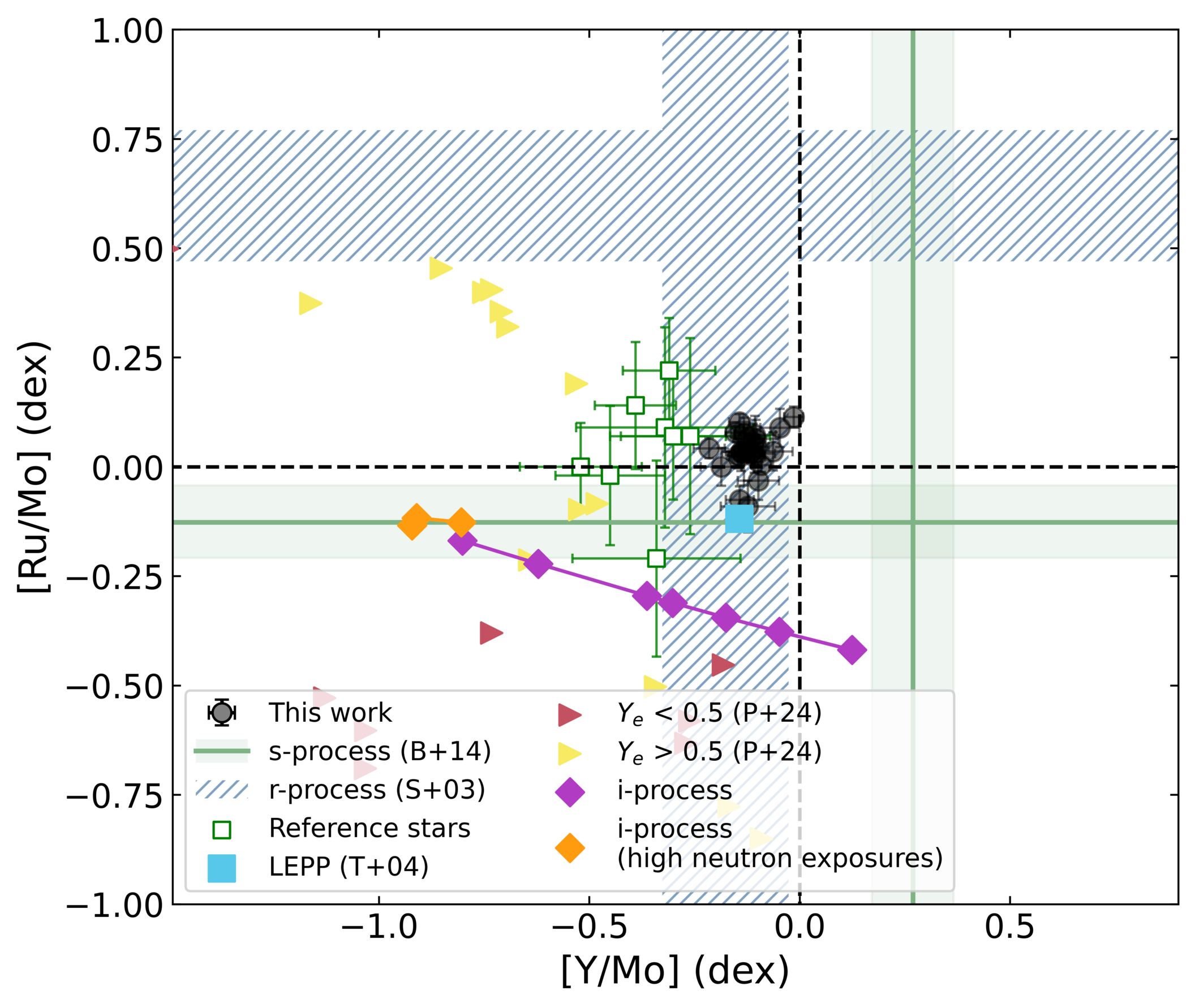}
      \caption{[Ru/Mo] for our sample of OCs with respect to [Zr/Mo], [Sr/Mo] and [Y/Mo] based on Fig.~3 from \citet{Mishenina2026}, whose results are shown as dark purple crosses. For comparison, the metal-poor stars from
\citet{Honda2006,Honda2007}, \citet{Hansen2012, Hansen2014} and \citet{Aoki2017} are shown as white squares with green contours.
The ratios from the s-process GCE simulations by \citet{Bisterzo2014}
(green line and error range) and from the star CS 22892-052
\citep{Sneden2003} (blue diagonal hashed lines) are included as references. We also report the LEPP
prediction by \citealt{Travaglio2004} (large cyan square), the ratios from
different neutrino-driven ejecta conditions in \citep{Psalti2024} with
$Y_e$ > 0.5 (yellow triangles, $\nu$p-process) and with $Y_e$ < 0.5 (red triangles,
weak r-process), along with i-process predictions for mild neutron
exposures \citet{Mishenina2026} shown as magenta-lined rotated squares and the i-process predictions for high neutron exposures shown as orange rotated squares. The dashed black lines that go through [0, 0] indicate the solar value. }
    \label{fig:chem_planes}
   \end{figure}

We investigate chemical planes to gain insight into the stellar sources and to place constraints on their nucleosynthetic origin. For this purpose, we follow the approach by \citet{Mishenina2026}, who analysed abundance patterns independently of metallicity by comparing [Zr/Mo] vs [Ru/Mo] ratios. We replicated this chemical plane in Fig.~\ref{fig:chem_planes} and extended the analysis by additionally exploring the [Sr/Mo] and [Y/Mo] ratios versus [Mo/Ru] for comparison. Consistent with \citet{Mishenina2026}, none of the stars exhibits lower [Ru/Mo] or higher [Zr/Mo] ratios than those predicted for the s-process by Galactic Chemical Evolution (GCE) models \citep{Bisterzo2014}. Across all three panels in Fig.~\ref{fig:chem_planes}, we find a systematic offset from predicted s-process ratios. This suggests that the s-process alone cannot account for the observed [Ru/Mo] values, as expected from GCE calculations for the MW disk and the Sun \citep[][]{Bisterzo2014, Prantzos2020}.
 
We exclude NLTE effects as the cause of this offset: our Sr and Y abundances are NLTE-corrected, and for Zr, Mo, and Ru, we utilised weak lines,  minimising NLTE impacts at the target metallicities.

The r-process contribution is usually estimated as the residual remaining after subtracting the s-process fraction (r- = 1 $-$ s-). Although this approach provides a useful reference, it implicitly assigns all yields not produced by the s-process to the r-process, neglecting the contributions from other nucleosynthetic channels, such as the p- and i-processes, which have been proposed as production channels for Mo and Ru (see Sect.~\ref{sec:nucleosynth}). For this reason, and following \citet{Mishenina2026},  we adopt the r-II star CS 22892-052 \citep{Sneden2003} as a representative of the r-process production in the Zr–Ru region, as it exhibits an abundance pattern consistent with the solar r-process residual for elements heavier than Ba \citep{Sneden2008}. While this star matches the solar r-process pattern for heavy elements (beyond Ba), this agreement breaks down for lighter elements between Sr and Ba \citep[see][and references therein]{Mishenina2026}. 

Such a discrepancy led to the introduction of the LEPP \citep[Lighter Element Primary Process, ][]{Travaglio2004,Montes2007}, an additional nucleosynthesis component required to reproduce solar abundances originally associated with the abundances observed in several metal-poor stars. The predicted LEPP component is also shown in Fig.~\ref{fig:chem_planes}, however, its contribution seems insufficient to explain the offset.

In the figure, we also include the different conditions of the neutrino-driven ejecta with varying electron fraction ($Y_e$) from \citet{Psalti2024}, noting that nucleosynthesis can proceed through either the weak r- ($Y_e$ < 0.5) or the $\nu$p-process ($Y_e$ > 0.5). Finally, we report the typical range of abundance ratios obtained under i-process conditions, calculated using the same trajectory reproducing i-process neutron density conditions used by \cite{Mishenina2026}, for neutron exposures yielding an efficient production of the Zr–Ru mass region without Ba production. There are no Ba abundances for stars in our sample, but it is still a matter of debate the presence of enhanced Ba in young OCs, whether due to an i-process contribution or as an observational artefact \citep[e.g., ][]{Mishenina2015, baratella:21}. However, for higher neutron exposures and higher Ba production, the i-process ratios are also shown in Fig.~\ref{fig:chem_planes} and are concentrated around the highest [Ru/Mo] values reported. 

From the comparison between our observations and the different nucleosynthesis components considered, we cannot clearly distinguish whether the location of our thin disc stars (OCs and field stars) might be reproduced by a simple mixture of s- and r-processes, or whether additional components would be required. This confirms the conclusion of \citet{Mishenina2026} for field stars concerning the observed Mo, Ru, and Zr abundance patterns. After the unexpected anomalies discovered in the Ba region \citep{Dorazi2009, Mishenina2015,baratella:21}, we cannot assume a perfect match between the signatures found in the field stars of the Galactic disc and OCs. Our results are based on a large, homogeneously analysed sample of OCs, which offer several advantages over field-star samples. Their members share a common age, kinematics, Galactic origin, and initial chemical composition, allowing properties such as age, distance, and chemical composition to be determined more accurately than for many other Galactic tracers \citep{Friel2013}, allowing us to further explore the chemical signatures in the Sr-Ru region.

\section{Summary and conclusions}\label{sec:conclu}

This work is based on high-resolution spectra from the SPA programme obtained with the HARPS-N echelle spectrograph at the TNG. Chemical abundances of Mo, Ru and Zr were derived through spectral synthesis using \texttt{TSFitPy} for 81 stars in 30 OCs. In particular, we provide the first Ru abundance measurements for a large sample of OCs.

To investigate the nucleosynthetic origins of Mo and Ru, we examine the relationships between their abundances and those of other neutron-capture elements (Sr, Y, Zr, and Eu). We find that Mo and Ru are tightly correlated, exhibiting slopes near unity when compared to each other and to s-process tracers (Sr, Y, and Zr). These results indicate similar enrichment timescales and are consistent with previous studies of thin-disc stars and OCs. In contrast, comparisons with the r-process element Eu show significant deviations from unity, consistent with the mixed nucleosynthetic origin and the corresponding isotopic composition. The strong correlation between Mo and Ru further supports their partially shared nucleosynthetic pathways. When studying the chemical planes of [Mo/Ru] versus [s/Mo] for the s-process elements Sr, Y and Zr, the resulting abundance patterns show systematic offsets with respect to pure s-process predictions from GCE models \citep{Bisterzo2014}, confirming that the s-process alone is insufficient to reproduce the observed ratios. These discrepancies cannot be attributed to NLTE effects or analysis systematics, as the abundances are either NLTE corrected or based on weak spectral lines.

While the derivation of a classical r-process residual for the solar system remains highly uncertain since it neglects contributions from other nucleosynthetic processes, this limitation further justifies the choice of using the r-II CS 22892-052 star as a representative r-process benchmark. Comparisons with this star and alternative nucleosynthetic scenarios further highlight the need for additional contributions beyond the classical s-process framework, although the relative contributions of different channels cannot be uniquely disentangled from the present data, preventing us from reliably favouring one specific combination or astrophysical source.

Overall, these results reinforce the view that Mo and Ru cannot be explained by a single dominant production site, but instead, originate from multiple nucleosynthetic processes operating on comparable timescales. As part of the future work on further constraining possible nucleosynthetic sources. Ba isotopic ratios can help distinguish between the relative contributions of the s-, r- and i-processes as isotopes are produced in different proportions \citep[see][and references therein]{Giribaldi2025, Giribaldi2026}. In particular, \citet{Liu2014},  showed that specific Ba isotopic signatures that could not be reproduced by standard AGB s-process calculations were consistent with nucleosynthesis under i-process conditions. In fact, the Ba isotopic ratios are becoming a target again for stars in different systems and at different metallicities \citep[see][]{HRMOS2026}. Therefore, future measurements of Ba isotopic ratios in our OC sample could provide an independent means of assessing the abundance patterns.

\begin{acknowledgements}
      
Based on observations made with the Italian Telescopio Nazionale Galileo (TNG) operated on the island of La Palma by the Fundación Galileo Galilei of the INAF (Istituto Nazionale di Astrofisica) at the Observatorio del Roque de los Muchachos. This research used facilities of the Italian Center for Astronomical Archive (IA2) operated by INAF at the Astronomical Observatory of Trieste. This work was done thanks to the support through an ERASMUS Mundus Joint Master scholarship Co-funded by the European Union under the call ERASMUS-EDU-2021-PEX-EMJM-MOB in the framework of the Erasmus+, Erasmus Mundus Joint Master in Astrophysics and Space Science – MASS. Views and opinions expressed are
however those of the author(s) only and do not necessarily reflect those
of the European Union or granting authority European Education and
Culture Executive Agency (EACEA). Neither the European Union nor
the granting authority can be held responsible for them.

This work has made use of data from the European Space Agency (ESA) mission Gaia (https://www.cosmos.esa.int/gaia), processed by the Gaia Data Processing and Analysis Consortium (DPAC, https://www.cosmos.esa.int/web/gaia/dpac/consortium). Funding for the DPAC has been provided by national institutions, in particular the institutions participating in the Gaia Multilateral
Agreement. This research has made use of the VizieR catalog access tool, CDS, Strasbourg, France (DOI : 10.26093/cds/vizier). The original description of the VizieR service was published in 2000, A\&AS 143, 23. Use of the NASA's Astrophysical Data System and TOPCAT \citep{topcat} are also acknowledged.

A.B. and V.D'O.  acknowledge funding from INAF MiniGrant 2022 (High resolution spectroscopy of open clusters). V.D'O. acknowledges support from INAF Minigrant 2024. 

A.P.  acknowledges the support of the Natural Sciences and Engineering Research Council of Canada (NSERC) under grant SAPIN-2026-00045.

T.M. and M.P. acknowledge ChETEC COST Action (CA16117), supported by COST (European Cooperation in Science and Technology), the ChETEC-INFRA project funded from the European Union's Horizon 2020 research and innovation programme (grant agreement No 101008324), and the IReNA network supported by NSF AccelNet (Grant No. OISE-1927130). 
M.P. also thanks the Lend\"ulet Program LP2023-10 of the Hungarian Academy of Sciences, the Hungarian NKFIH via K-project 138031 and NKKP Advanced grant 153697, the 
support from the ERC Synergy Grant Programme (Geoastronomy, grant agreement number 101166936, Germany), and the use of the Astrohub online virtual research environment (https://astrohub.uvic.ca), developed and operated by the Computational Stellar Astrophysics group (http://csa.phys.uvic.ca) at the University of Victoria and hosted on the Computed Canada Arbutus Cloud at the University of Victoria.
T.M. acknowledges support from Grant PID2023-147569NB-C21 funded by MICIU/AEI/10.13039/501100011033 Spain and ERDF, UE.
A.N. acknowledges that part of the research activities described in this report were
carried out with contribution of the Next Generation EU funds within the
National Recovery and Resilience Plan (PNRR), Mission 4 - Education and Research, Component 2 - From Research to Business (M4C2), Investment Line 3.1 - Strengthening and creation of Research Infrastructures, Project
IR0000034 – “STILES - Strengthening the Italian Leadership in ELT and SKA”, CUP C33C22000640006. 
\end{acknowledgements}

\bibliographystyle{aa} 
\bibliography{mo_ru_ocs}

\begin{appendix}

\onecolumn

    \begin{figure*}
    \section{Colour-magnitude diagrams (CMDs) for the studied clusters}
        \centering
        \includegraphics[width=0.83\textwidth]{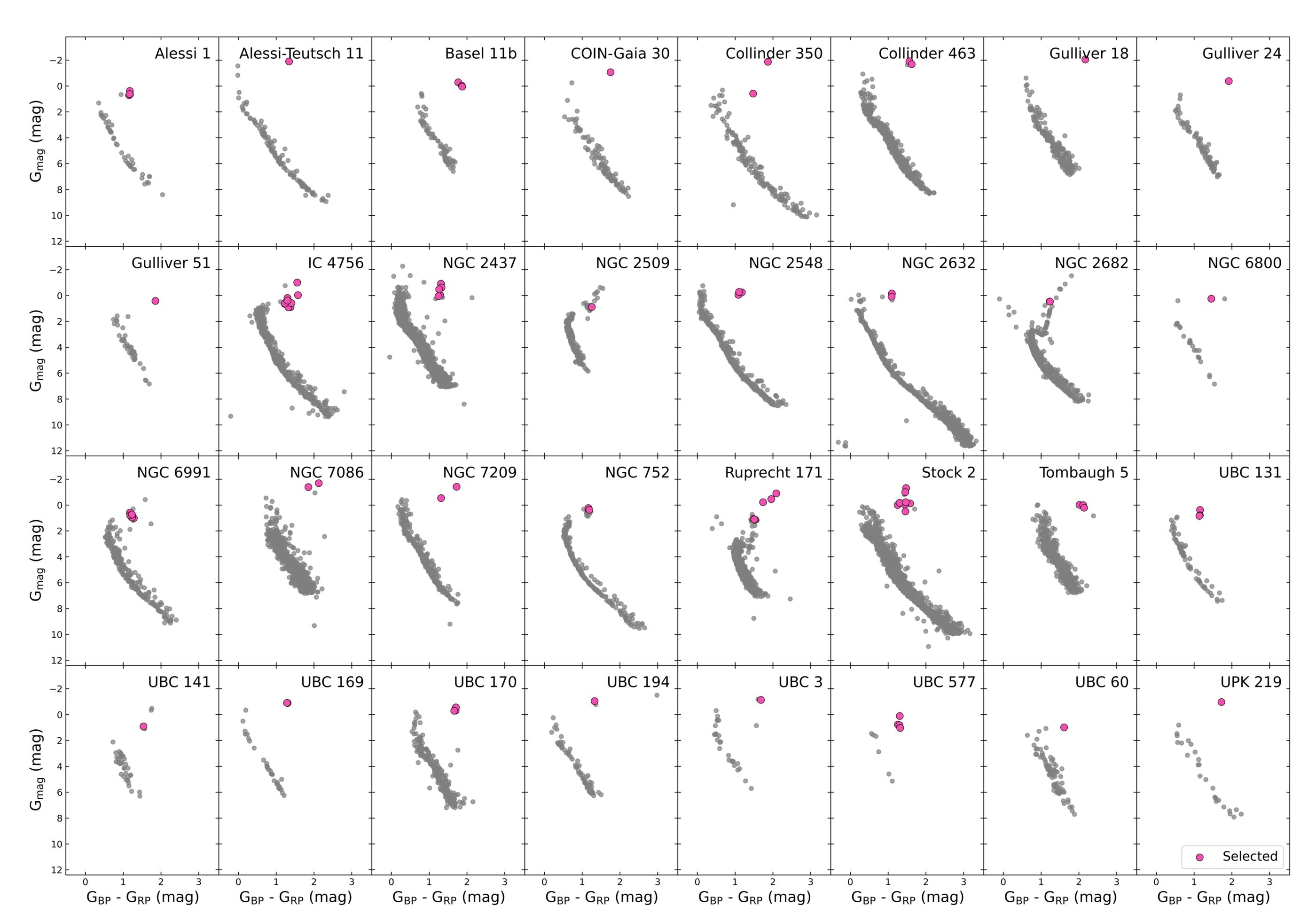}
        \caption{CMDs of the studied OCs using the members determined by \citet{Cantat-Gaudin2020}.}
        \label{fig:cmds}%
    \end{figure*}

\section{Stellar parameters for the studied members}

\begin{longtable}{lcccccc}
\caption{Stellar parameters for 81 stars considered in this work.}\\
\label{table:stellarparams} \\
\hline\hline
Star & Gaia DR3 ID & T$_{\text{eff}}$ \tablefootmark{a} & $\log g$ \tablefootmark{a} & $v_{\text{mic}}$ \tablefootmark{a} & $v_{\text{mac}}$ \tablefootmark{b} & [Fe/H] \tablefootmark{a}\\
 & & (K) & (dex) & kms$^{-1}$ & kms$^{-1}$ & (dex) \\
\hline
\endfirsthead
\caption{continued.}\\
\hline
Star & Gaia DR3 ID & T$_{\text{eff}}$ \tablefootmark{a} & $\log g$ \tablefootmark{a} & $v_{\text{mic}}$ \tablefootmark{a} & $v_{\text{mac}}$ \tablefootmark{b} & [Fe/H] \tablefootmark{a}\\
 & & (K) & (dex) & kms$^{-1}$ & kms$^{-1}$ & (dex) \\
\hline
\endhead
\hline
\endfoot
Alessi 161 star \#1 & 4505874693061489280 & $4731 \pm 32$ & $2.32 \pm 0.09$ & $1.84 \pm 0.01$ & $4.16 \pm 0.21$ & $0.00 \pm 0.01$ \\
Alessi 1 star \#2 & 402506369136008832 & $4986 \pm 30$ & $2.83 \pm 0.07$ & $1.30 \pm 0.01$ & $3.22 \pm 0.06$ & $-0.00 \pm 0.01$ \\
Alessi 1 star \#3 & 402505991180022528 & $4996 \pm 30$ & $2.84 \pm 0.07$ & $1.28 \pm 0.01$ & $3.11 \pm 0.08$ & $-0.01 \pm 0.01$ \\
Alessi 1 star \#5 & 402867593065772288 & $4939 \pm 31$ & $2.73 \pm 0.08$ & $1.35 \pm 0.01$ & $3.29 \pm 0.07$ & $-0.03 \pm 0.01$ \\
Alessi 1 star \#6 & 402880684126058880 & $4985 \pm 32$ & $2.87 \pm 0.08$ & $1.28 \pm 0.01$ & $2.85 \pm 0.03$ & $-0.01 \pm 0.01$ \\
Alessi-Teutsch 11 star \#1 & 2184332753719499904 & $4517 \pm 35$ & $2.15 \pm 0.11$ & $1.76 \pm 0.01$ & $3.74 \pm 0.14$ & $-0.04 \pm 0.01$ \\
Basel11b star1 & 3424056131485038592 & $4950 \pm 35$ & $2.51 \pm 0.09$ & $1.81 \pm 0.01$ & $5.05 \pm 0.17$ & $0.01 \pm 0.01$ \\
Basel11b star2 & 3424055921028900736 & $5008 \pm 33$ & $2.64 \pm 0.08$ & $1.77 \pm 0.01$ & $4.45 \pm 0.13$ & $0.02 \pm 0.01$ \\
Basel11b star3 & 3424057540234289408 & $4760 \pm 34$ & $2.55 \pm 0.10$ & $1.47 \pm 0.02$ & $4.21 \pm 0.14$ & $0.08 \pm 0.01$ \\
COIN-Gaia30 star1 & 532533682228608384 & $4796 \pm 35$ & $2.41 \pm 0.10$ & $1.69 \pm 0.01$ & $4.72 \pm 0.16$ & $-0.04 \pm 0.01$ \\
Collinder 350 star \#2 & 4372572888274176768 & $5183 \pm 33$ & $3.06 \pm 0.07$ & $1.35 \pm 0.02$ & $6.25 \pm 0.18$ & $0.08 \pm 0.01$ \\
Collinder 463 star \#1 & 534207555539397888 & $4557 \pm 35$ & $2.01 \pm 0.11$ & $1.96 \pm 0.01$ & $3.66 \pm 0.21$ & $-0.04 \pm 0.01$ \\
Collinder 463 star \#3 & 534363067715447680 & $4634 \pm 33$ & $2.00 \pm 0.11$ & $2.07 \pm 0.01$ & $4.18 \pm 0.19$ & $-0.07 \pm 0.01$ \\
Gulliver 51 star \#1 & 517925575042048384 & $4574 \pm 36$ & $2.14 \pm 0.12$ & $1.53 \pm 0.01$ & $2.94 \pm 0.18$ & $-0.17 \pm 0.01$ \\
IC 4756 star \#0 & 4283940671842998272 & $5048 \pm 30$ & $2.91 \pm 0.07$ & $1.34 \pm 0.01$ & $3.31 \pm 0.04$ & $-0.04 \pm 0.01$ \\
IC 4756 star \#2 & 4283984931511649792 & $4451 \pm 33$ & $2.09 \pm 0.10$ & $1.52 \pm 0.02$ & $3.33 \pm 0.43$ & $-0.06 \pm 0.01$ \\
IC 4756 star \#4 & 4283901746580386816 & $4580 \pm 31$ & $2.34 \pm 0.09$ & $1.28 \pm 0.02$ & $3.15 \pm 0.37$ & $-0.02 \pm 0.01$ \\
IC 4756 star \#5 & 4283961979205354496 & $5053 \pm 34$ & $3.01 \pm 0.08$ & $1.26 \pm 0.02$ & $2.83 \pm 0.07$ & $0.06 \pm 0.01$ \\
IC 4756 star \#7 & 4284658038773033728 & $5026 \pm 29$ & $2.85 \pm 0.07$ & $1.32 \pm 0.01$ & $2.91 \pm 0.06$ & $-0.04 \pm 0.01$ \\
IC 4756 star \#8 & 4283900062953084800 & $5074 \pm 34$ & $2.99 \pm 0.08$ & $1.30 \pm 0.02$ & $3.08 \pm 0.11$ & $0.04 \pm 0.01$ \\
IC 4756 star \#9 & 4284802074805175936 & $5161 \pm 34$ & $3.12 \pm 0.08$ & $1.26 \pm 0.02$ & $3.26 \pm 0.06$ & $0.00 \pm 0.01$ \\
IC 4756 star \#11 & 4283789905627894272 & $5043 \pm 29$ & $2.93 \pm 0.07$ & $1.30 \pm 0.01$ & $3.25 \pm 0.07$ & $-0.05 \pm 0.01$ \\
IC 4756 star \#12 & 4283983552796484864 & $5037 \pm 30$ & $2.95 \pm 0.07$ & $1.25 \pm 0.02$ & $3.65 \pm 0.04$ & $-0.02 \pm 0.01$ \\
IC 4756 star \#13 & 4284806438475643776 & $5175 \pm 32$ & $3.14 \pm 0.07$ & $1.17 \pm 0.02$ & $3.33 \pm 0.07$ & $0.03 \pm 0.01$ \\
IC 4756 star \#14 & 4283997575895463680 & $5152 \pm 34$ & $3.16 \pm 0.07$ & $1.15 \pm 0.02$ & $2.84 \pm 0.08$ & $0.07 \pm 0.01$ \\
IC 4756 star \#15 & 4286303114338968448 & $4998 \pm 34$ & $2.88 \pm 0.08$ & $1.32 \pm 0.02$ & $3.11 \pm 0.37$ & $0.03 \pm 0.01$ \\
LP 1800 star \#1 & 2007935075311695872 & $4810 \pm 42$ & $2.53 \pm 0.12$ & $1.73 \pm 0.02$ & $6.13 \pm 0.44$ & $0.12 \pm 0.01$ \\
LP 1800 star \#2 & 2007981499630067456 & $4979 \pm 40$ & $2.65 \pm 0.10$ & $1.87 \pm 0.01$ & $8.09 \pm 0.08$ & $0.08 \pm 0.01$ \\
LP 1800 star \#3 & 2007934044519547904 & $4904 \pm 36$ & $2.54 \pm 0.09$ & $1.69 \pm 0.01$ & $3.83 \pm 0.15$ & $0.08 \pm 0.01$ \\
NGC 2437 star \#1 & 3029609393042459392 & $4911 \pm 36$ & $2.56 \pm 0.10$ & $1.73 \pm 0.01$ & $4.73 \pm 0.15$ & $0.08 \pm 0.01$ \\
NGC 2437 star \#10 & 3029226694277998080 & $4513 \pm 49$ & $1.77 \pm 0.16$ & $1.05 \pm 0.02$ & $4.30 \pm 0.72$ & $-0.18 \pm 0.02$ \\
NGC 2437 star \#3 & 3030364134752459904 & $5010 \pm 40$ & $2.66 \pm 0.11$ & $1.80 \pm 0.01$ & $8.64 \pm 0.37$ & $-0.03 \pm 0.01$ \\
NGC 2437 star \#4 & 3029132686034894592 & $4835 \pm 31$ & $2.49 \pm 0.09$ & $1.51 \pm 0.01$ & $3.12 \pm 0.14$ & $-0.06 \pm 0.01$ \\
NGC 2437 star \#5 & 3029156222454419072 & $4924 \pm 35$ & $2.74 \pm 0.09$ & $1.37 \pm 0.02$ & $3.00 \pm 0.16$ & $0.04 \pm 0.01$ \\
NGC 2509 star \#18 & 5714209934411718784 & $4773 \pm 40$ & $2.89 \pm 0.10$ & $1.06 \pm 0.04$ & $2.92 \pm 0.25$ & $0.21 \pm 0.01$ \\
NGC 2548 star \#1 & 3064481400744808704 & $5102 \pm 35$ & $3.01 \pm 0.08$ & $1.48 \pm 0.02$ & $8.69 \pm 0.37$ & $0.07 \pm 0.01$ \\
NGC 2548 star \#2 & 3064537647636773760 & $5063 \pm 32$ & $2.90 \pm 0.08$ & $1.40 \pm 0.02$ & $3.69 \pm 0.06$ & $0.10 \pm 0.01$ \\
NGC 2548 star \#3 & 3064579703955646976 & $4715 \pm 33$ & $2.43 \pm 0.09$ & $1.40 \pm 0.01$ & $2.81 \pm 0.05$ & $-0.03 \pm 0.01$ \\
NGC2632 HD73665 & 661324431287688448 & $4846 \pm 34$ & $2.58 \pm 0.09$ & $1.45 \pm 0.02$ & $2.90 \pm 0.08$ & $0.13 \pm 0.01$ \\
NGC 2632 star \#1 & 661271173693364864 & $4911 \pm 35$ & $2.84 \pm 0.09$ & $1.26 \pm 0.03$ & $2.97 \pm 0.05$ & $0.20 \pm 0.01$ \\
NGC 2682 star \#8 & 604921512005266048 & $4726 \pm 33$ & $2.62 \pm 0.09$ & $1.25 \pm 0.02$ & $3.15 \pm 0.36$ & $0.06 \pm 0.01$ \\
NGC 2682 star \#9 & 604920202039656064 & $4718 \pm 33$ & $2.59 \pm 0.09$ & $1.27 \pm 0.02$ & $3.02 \pm 0.39$ & $0.02 \pm 0.01$ \\
NGC 6800 star \#1 & 2023140603917404416 & $5051 \pm 33$ & $2.87 \pm 0.08$ & $1.40 \pm 0.02$ & $3.66 \pm 0.20$ & $0.11 \pm 0.01$ \\
NGC 6991 star \#4 & 2166845227448342400 & $5017 \pm 32$ & $3.03 \pm 0.08$ & $1.21 \pm 0.02$ & $3.12 \pm 0.07$ & $0.03 \pm 0.01$ \\
NGC 6991 star \#6 & 2167008230043576192 & $4853 \pm 30$ & $2.74 \pm 0.08$ & $1.24 \pm 0.02$ & $2.96 \pm 0.06$ & $-0.02 \pm 0.01$ \\
NGC 6991 star \#7 & 2166853228954093312 & $4994 \pm 29$ & $2.93 \pm 0.07$ & $1.25 \pm 0.01$ & $3.09 \pm 0.13$ & $-0.03 \pm 0.01$ \\
NGC 6991 star \#9 & 2166908105771609472 & $5084 \pm 36$ & $3.14 \pm 0.08$ & $1.11 \pm 0.03$ & $2.67 \pm 0.37$ & $0.11 \pm 0.01$ \\
NGC 6991 star \#10 & 2166821789793563904 & $5078 \pm 36$ & $3.12 \pm 0.08$ & $1.16 \pm 0.03$ & $2.91 \pm 0.05$ & $0.09 \pm 0.01$ \\
NGC 7086 star \#1 & 2171663253032105984 & $4545 \pm 37$ & $2.03 \pm 0.12$ & $1.93 \pm 0.01$ & $3.70 \pm 0.24$ & $-0.05 \pm 0.01$ \\
NGC 7086 star \#2 & 2171707542735344000 & $4471 \pm 37$ & $1.79 \pm 0.13$ & $2.06 \pm 0.01$ & $4.34 \pm 0.27$ & $-0.11 \pm 0.01$ \\
NGC 7209 star \#2 & 1975004019170020736 & $4873 \pm 35$ & $2.52 \pm 0.10$ & $1.65 \pm 0.01$ & $4.19 \pm 0.17$ & $0.04 \pm 0.01$ \\
NGC 752 star \#3 & 342937195667536512 & $4931 \pm 33$ & $2.92 \pm 0.08$ & $1.21 \pm 0.02$ & $2.88 \pm 0.08$ & $0.07 \pm 0.01$ \\
NGC 752 star \#4 & 342899537393760512 & $4920 \pm 31$ & $2.88 \pm 0.08$ & $1.23 \pm 0.02$ & $2.90 \pm 0.08$ & $0.06 \pm 0.01$ \\
NGC 752 star \#5 & 342890127122193280 & $4887 \pm 34$ & $2.90 \pm 0.08$ & $1.13 \pm 0.03$ & $2.74 \pm 0.17$ & $0.10 \pm 0.01$ \\
Ruprecht 171 star \#4 & 4103101073850814208 & $4593 \pm 36$ & $2.28 \pm 0.11$ & $1.32 \pm 0.01$ & $2.65 \pm 0.14$ & $-0.08 \pm 0.01$ \\
Ruprecht 171 star \#5 & 4103072727066663552 & $4630 \pm 35$ & $2.32 \pm 0.11$ & $1.37 \pm 0.01$ & $2.74 \pm 0.17$ & $-0.09 \pm 0.01$ \\
Ruprecht 171 star \#6 & 4103098050193837952 & $4606 \pm 35$ & $2.34 \pm 0.11$ & $1.29 \pm 0.01$ & $2.67 \pm 0.12$ & $-0.08 \pm 0.01$ \\
Ruprecht 171 star \#8 & 4102884023383492096 & $4617 \pm 35$ & $2.45 \pm 0.10$ & $1.23 \pm 0.03$ & $2.53 \pm 0.08$ & $0.02 \pm 0.01$ \\
Stock 2 star \#1 & 465132764751065984 & $4932 \pm 31$ & $2.61 \pm 0.08$ & $1.56 \pm 0.01$ & $3.56 \pm 0.08$ & $0.01 \pm 0.01$ \\
Stock 2 star \#2 & 458067680993514880 & $4402 \pm 38$ & $1.89 \pm 0.13$ & $1.58 \pm 0.02$ & $3.27 \pm 0.19$ & $-0.14 \pm 0.01$ \\
Stock 2 star \#4 & 507214579443494144 & $4984 \pm 31$ & $2.70 \pm 0.08$ & $1.50 \pm 0.01$ & $3.14 \pm 0.09$ & $0.05 \pm 0.01$ \\
Stock 2 star \#6 & 507520106232760320 & $5088 \pm 28$ & $2.85 \pm 0.07$ & $1.31 \pm 0.01$ & $4.61 \pm 0.14$ & $-0.02 \pm 0.01$ \\
Stock 2 star \#7 & 507507702367267584 & $4852 \pm 31$ & $2.46 \pm 0.09$ & $1.54 \pm 0.01$ & $3.31 \pm 0.16$ & $-0.04 \pm 0.01$ \\
Stock 2 star \#8 & 459118882826608640 & $5084 \pm 29$ & $2.83 \pm 0.07$ & $1.39 \pm 0.01$ & $5.13 \pm 0.13$ & $-0.02 \pm 0.01$ \\
Stock 2 star \#9 & 459112148318029056 & $5038 \pm 30$ & $2.82 \pm 0.07$ & $1.42 \pm 0.01$ & $5.25 \pm 0.13$ & $0.04 \pm 0.01$ \\
Stock 2 star \#10 & 459199662573391104 & $4800 \pm 36$ & $2.53 \pm 0.10$ & $1.54 \pm 0.02$ & $6.59 \pm 0.30$ & $-0.05 \pm 0.01$ \\
Tombaugh 5 star \#2 & 473275782228263296 & $4800 \pm 46$ & $2.35 \pm 0.14$ & $1.98 \pm 0.02$ & $3.68 \pm 0.11$ & $-0.08 \pm 0.02$ \\
Tombaugh 5 star \#3 & 473268424940932864 & $4932 \pm 36$ & $2.51 \pm 0.10$ & $1.88 \pm 0.01$ & $3.95 \pm 0.18$ & $0.07 \pm 0.01$ \\
Tombaugh 5 star \#4 & 473266779976916480 & $4736 \pm 43$ & $2.37 \pm 0.13$ & $1.76 \pm 0.02$ & $5.52 \pm 0.37$ & $0.06 \pm 0.01$ \\
UBC 141 star \#1 & 1869885794120648192 & $4643 \pm 32$ & $2.61 \pm 0.09$ & $1.13 \pm 0.03$ & $2.50 \pm 0.37$ & $-0.02 \pm 0.01$ \\
UBC 169 star \#2 & 1988520899724392064 & $4995 \pm 45$ & $2.67 \pm 0.12$ & $1.87 \pm 0.02$ & $8.39 \pm 0.16$ & $0.08 \pm 0.01$ \\
UBC 194 star \#1 & 354343808469156608 & $5057 \pm 48$ & $2.76 \pm 0.12$ & $1.97 \pm 0.01$ & $9.34 \pm 0.32$ & $0.06 \pm 0.01$ \\
UBC 577 star \#1 & 4531532690223665664 & $5155 \pm 37$ & $3.21 \pm 0.08$ & $1.22 \pm 0.03$ & $2.81 \pm 0.12$ & $0.02 \pm 0.01$ \\
UBC 577 star \#2 & 4531526058785223424 & $5040 \pm 31$ & $3.02 \pm 0.07$ & $1.11 \pm 0.02$ & $2.72 \pm 0.07$ & $-0.06 \pm 0.01$ \\
UBC 577 star \#3 & 4531525337240008576 & $5051 \pm 29$ & $2.94 \pm 0.07$ & $1.21 \pm 0.01$ & $3.33 \pm 0.17$ & $-0.07 \pm 0.01$ \\
UBC 577 star \#5 & 4531474038155634560 & $5109 \pm 33$ & $3.10 \pm 0.08$ & $1.21 \pm 0.02$ & $3.04 \pm 0.12$ & $-0.02 \pm 0.01$ \\
UBC 60 star \#3 & 179623851673246464 & $4990 \pm 35$ & $2.94 \pm 0.08$ & $1.25 \pm 0.03$ & $3.44 \pm 0.04$ & $0.23 \pm 0.01$ \\
UPK 219 star \#1 & 2209440823287736064 & $4913 \pm 37$ & $2.47 \pm 0.10$ & $1.93 \pm 0.01$ & $4.88 \pm 0.15$ & $0.07 \pm 0.01$ \\
UPK 84 star \#1 & 1814336194629117824 & $5066 \pm 34$ & $3.01 \pm 0.08$ & $1.27 \pm 0.02$ & $2.93 \pm 0.05$ & $0.01 \pm 0.01$ \\
UPK 84 star \#2 & 1817359031271791232 & $5106 \pm 34$ & $3.17 \pm 0.08$ & $1.15 \pm 0.02$ & $2.94 \pm 0.04$ & $0.06 \pm 0.01$ \\
UPK 84 star \#3 & 1814243324558277248 & $5120 \pm 35$ & $3.22 \pm 0.08$ & $1.06 \pm 0.03$ & $3.21 \pm 0.03$ & $0.08 \pm 0.01$ \\
\end{longtable}
\tablefoot{\tablefoottext{a}{Obtained by \citetalias{SPA_DalPonte2025}.} \tablefoottext{b}{Estimated for this work using the results from \citetalias{SPA_DalPonte2025}.}}

\section{Estimated abundances star by star}

\begin{longtable}{lccccc}
\caption{Estimated abundances for Mo, Ru and Zr for each star in our sample, together with the number of spectral lines used to derive the mean abundances of Mo and Zr. The number of lines for Ru is not reported, as it is based on a single line.}\\
\label{table:Abund_stars} \\
\hline\hline
Star & [Mo/Fe] & $N_{\mathrm{lines}}^{\mathrm{Mo}}$  & [Ru/Fe] & [Zr/Fe] & $N_{\mathrm{lines}}^{\mathrm{Zr}}$ \\
 & (dex) & & (dex) &(dex) & \\
\hline
\endfirsthead
\caption{continued.}\\
\hline
Star & [Mo/Fe] & $N_{\mathrm{lines}}^{\mathrm{Mo}}$  & [Ru/Fe] & [Zr/Fe] & $N_{\mathrm{lines}}^{\mathrm{Zr}}$ \\
 & (dex) & & (dex) &(dex) & \\
\hline
\endhead
\hline
\endfoot
Alessi 161 star \#1 & $0.06 \pm 0.01$ & 2 & $0.13 \pm 0.04$ & $-0.04 \pm 0.03$ & 2 \\
Alessi 1 star \#2 & $0.06 \pm 0.04$ & 1 & $0.16 \pm 0.05$ & $-0.06 \pm 0.01$ & 2 \\
Alessi 1 star \#3 & $0.14 \pm 0.04$ & 1 & $0.04 \pm 0.05$ & $-0.03 \pm 0.02$ & 2 \\
Alessi 1 star \#5 & $0.06 \pm 0.01$ & 2 & $0.09 \pm 0.05$ & $-0.08 \pm 0.02$ & 2 \\
Alessi 1 star \#6 & $0.05 \pm 0.04$ & 1 & $0.11 \pm 0.05$ & $-0.08 \pm 0.01$ & 2 \\
Alessi-Teutsch 11 star \#1 & $0.02 \pm 0.04$ & 2 & $0.05 \pm 0.04$ & $-0.13 \pm 0.02$ & 2 \\
Basel11b star1 & $0.07 \pm 0.04$ & 1 & $-$ & $0.00 \pm 0.01$ & 2 \\
Basel11b star2 & $0.11 \pm 0.04$ & 1 & $-$ & $0.08 \pm 0.03$ & 2 \\
Basel11b star3 & $0.01 \pm 0.04$ & 1 & $-$ & $-0.06 \pm 0.02$ & 2 \\
COIN-Gaia30 star1 & $0.05 \pm 0.04$ & 1 & $0.02 \pm 0.04$ & $0.00 \pm 0.01$ & 2 \\
Collinder 350 star \#2 & $0.12 \pm 0.04$ & 1 & $-$ & $0.07 \pm 0.02$ & 2 \\
Collinder 463 star \#1 & $0.04 \pm 0.04$ & 2 & $0.12 \pm 0.03$ & $-0.12 \pm 0.02$ & 2 \\
Collinder 463 star \#3 & $0.08 \pm 0.06$ & 2 & $0.15 \pm 0.03$ & $-0.06 \pm 0.02$ & 2 \\
Gulliver 51 star \#1 & $-0.04 \pm 0.03$ & 2 & $0.02 \pm 0.04$ & $-0.14 \pm 0.03$ & 2 \\
IC 4756 star \#0 & $0.08 \pm 0.07$ & 1 & $0.05 \pm 0.05$ & $0.02 \pm 0.01$ & 2 \\
IC 4756 star \#2 & $0.04 \pm 0.03$ & 2 & $0.09 \pm 0.05$ & $-0.08 \pm 0.03$ & 2 \\
IC 4756 star \#4 & $-0.01 \pm 0.07$ & 1 & $0.05 \pm 0.05$ & $-0.10 \pm 0.03$ & 2 \\
IC 4756 star \#5 & $0.10 \pm 0.01$ & 2 & $0.17 \pm 0.05$ & $0.03 \pm 0.01$ & 2 \\
IC 4756 star \#7 & $0.06 \pm 0.07$ & 1 & $0.13 \pm 0.05$ & $-0.07 \pm 0.01$ & 2 \\
IC 4756 star \#8 & $0.12 \pm 0.01$ & 2 & $0.16 \pm 0.05$ & $0.03 \pm 0.02$ & 2 \\
IC 4756 star \#9 & $0.20 \pm 0.07$ & 1 & $-$ & $0.03 \pm 0.01$ & 2 \\
IC 4756 star \#11 & $0.19 \pm 0.01$ & 2 & $0.22 \pm 0.05$ & $0.00 \pm 0.01$ & 2 \\
IC 4756 star \#12 & $0.16 \pm 0.07$ & 1 & $0.16 \pm 0.05$ & $0.01 \pm 0.01$ & 2 \\
IC 4756 star \#13 & $0.14 \pm 0.07$ & 1 & $0.12 \pm 0.05$ & $0.10 \pm 0.01$ & 2 \\
IC 4756 star \#14 & $0.18 \pm 0.07$ & 1 & $0.15 \pm 0.05$ & $0.07 \pm 0.02$ & 2 \\
IC 4756 star \#15 & $0.09 \pm 0.07$ & 1 & $0.18 \pm 0.05$ & $0.01 \pm 0.01$ & 2 \\
LP 1800 star \#1 & $0.01 \pm 0.01$ & 2 & $0.03 \pm 0.02$ & $-0.10 \pm 0.01$ & 2 \\
LP 1800 star \#2 & $0.05 \pm 0.02$ & 1 & $-$ & $-0.05 \pm 0.01$ & 2 \\
LP 1800 star \#3 & $0.07 \pm 0.01$ & 2 & $0.06 \pm 0.02$ & $-0.04 \pm 0.02$ & 2 \\
NGC 2437 star \#1 & $0.14 \pm 0.05$ & 2 & $0.09 \pm 0.03$ & $0.01 \pm 0.01$ & 2 \\
NGC 2437 star \#3 & $0.12 \pm 0.06$ & 1 & $-$ & $0.04 \pm 0.01$ & 2 \\
NGC 2437 star \#4 & $0.09 \pm 0.06$ & 2 & $0.14 \pm 0.03$ & $-0.09 \pm 0.01$ & 2 \\
NGC 2437 star \#5 & $0.11 \pm 0.06$ & 1 & $0.12 \pm 0.03$ & $-0.02 \pm 0.01$ & 2 \\
NGC 2437 star \#10 & $-0.04 \pm 0.03$ & 2 & $0.15 \pm 0.03$ & $-0.19 \pm 0.02$ & 2 \\
NGC 2509 star \#18 & $-0.09 \pm 0.04$ & 1 & $-$ & $-0.14 \pm 0.02$ & 1 \\
NGC 2548 star \#1 & $0.06 \pm 0.08$ & 1 & $-$ & $0.01 \pm 0.02$ & 2 \\
NGC 2548 star \#2 & $0.18 \pm 0.07$ & 2 & $0.21 \pm 0.14$ & $0.02 \pm 0.02$ & 2 \\
NGC 2548 star \#3 & $-0.02 \pm 0.01$ & 2 & $0.01 \pm 0.14$ & $-0.17 \pm 0.03$ & 2 \\
NGC2632 HD73665 & $-0.05 \pm 0.02$ & 2 & $0.05 \pm 0.03$ & $-0.21 \pm 0.01$ & 2 \\
NGC 2632 star \#1 & $-0.01 \pm 0.01$ & 2 & $0.09 \pm 0.03$ & $-0.14 \pm 0.01$ & 2 \\
NGC 2682 star \#8 & $0.01 \pm 0.01$ & 2 & $0.06 \pm 0.01$ & $-0.15 \pm 0.02$ & 2 \\
NGC 2682 star \#9 & $-0.03 \pm 0.02$ & 2 & $0.08 \pm 0.01$ & $-0.14 \pm 0.01$ & 2 \\
NGC 6800 star \#1 & $0.16 \pm 0.04$ & 1 & $0.22 \pm 0.04$ & $0.11 \pm 0.01$ & 2 \\
NGC 6991 star \#4 & $0.13 \pm 0.05$ & 1 & $0.12 \pm 0.04$ & $0.05 \pm 0.01$ & 2 \\
NGC 6991 star \#6 & $0.02 \pm 0.07$ & 2 & $0.10 \pm 0.04$ & $-0.08 \pm 0.01$ & 2 \\
NGC 6991 star \#7 & $0.19 \pm 0.05$ & 1 & $0.13 \pm 0.04$ & $-0.01 \pm 0.01$ & 2 \\
NGC 6991 star \#9 & $0.13 \pm 0.05$ & 1 & $0.20 \pm 0.04$ & $0.01 \pm 0.01$ & 2 \\
NGC 6991 star \#10 & $0.09 \pm 0.06$ & 2 & $0.16 \pm 0.04$ & $0.00 \pm 0.03$ & 2 \\
NGC 7086 star \#1 & $0.03 \pm 0.06$ & 2 & $0.08 \pm 0.02$ & $-0.11 \pm 0.02$ & 2 \\
NGC 7086 star \#2 & $0.02 \pm 0.04$ & 2 & $0.06 \pm 0.02$ & $-0.10 \pm 0.02$ & 2 \\
NGC 7209 star \#2 & $-0.01 \pm 0.03$ & 2 & $0.02 \pm 0.04$ & $-0.11 \pm 0.01$ & 2 \\
NGC 752 star \#3 & $-0.01 \pm 0.03$ & 1 & $0.16 \pm 0.05$ & $-0.03 \pm 0.02$ & 2 \\
NGC 752 star \#4 & $0.05 \pm 0.03$ & 1 & $0.14 \pm 0.05$ & $-0.04 \pm 0.01$ & 2 \\
NGC 752 star \#5 & $0.00 \pm 0.03$ & 1 & $0.07 \pm 0.05$ & $-0.06 \pm 0.02$ & 2 \\
Ruprecht 171 star \#4 & $-0.08 \pm 0.03$ & 1 & $-0.07 \pm 0.09$ & $-0.30 \pm 0.03$ & 2 \\
Ruprecht 171 star \#5 & $-0.06 \pm 0.03$ & 1 & $0.04 \pm 0.09$ & $-0.23 \pm 0.03$ & 2 \\
Ruprecht 171 star \#6 & $-0.13 \pm 0.03$ & 1 & $-0.18 \pm 0.09$ & $-0.33 \pm 0.03$ & 2 \\
Ruprecht 171 star \#8 & $-0.09 \pm 0.03$ & 1 & $-0.04 \pm 0.09$ & $-0.27 \pm 0.03$ & 2 \\
Stock 2 star \#1 & $0.07 \pm 0.01$ & 2 & $0.14 \pm 0.04$ & $-0.07 \pm 0.01$ & 2 \\
Stock 2 star \#2 & $0.00 \pm 0.05$ & 2 & $0.05 \pm 0.04$ & $-0.16 \pm 0.03$ & 2 \\
Stock 2 star \#4 & $0.10 \pm 0.01$ & 2 & $0.15 \pm 0.04$ & $0.00 \pm 0.01$ & 2 \\
Stock 2 star \#6 & $0.15 \pm 0.02$ & 2 & $-$ & $0.06 \pm 0.01$ & 2 \\
Stock 2 star \#7 & $0.00 \pm 0.03$ & 2 & $0.14 \pm 0.04$ & $-0.09 \pm 0.01$ & 2 \\
Stock 2 star \#8 & $0.13 \pm 0.02$ & 2 & $-$ & $0.03 \pm 0.02$ & 2 \\
Stock 2 star \#9 & $0.12 \pm 0.02$ & 2 & $0.15 \pm 0.04$ & $0.04 \pm 0.03$ & 2 \\
Stock 2 star \#10 & $0.21 \pm 0.03$ & 2 & $0.16 \pm 0.04$ & $0.13 \pm 0.02$ & 2 \\
Tombaugh 5 star \#2 & $0.16 \pm 0.12$ & 1 & $-$ & $0.04 \pm 0.01$ & 2 \\
Tombaugh 5 star \#3 & $0.21 \pm 0.12$ & 1 & $-$ & $0.01 \pm 0.02$ & 2 \\
Tombaugh 5 star \#4 & $-0.05 \pm 0.12$ & 1 & $0.02 \pm 0.04$ & $-0.14 \pm 0.05$ & 2 \\
UBC 141 star \#1 & $0.00 \pm 0.04$ & 1 & $0.04 \pm 0.04$ & $-0.15 \pm 0.03$ & 2 \\
UBC 169 star \#2 & $0.08 \pm 0.04$ & 1 & $-$ & $0.07 \pm 0.02$ & 2 \\
UBC 194 star \#1 & $0.02 \pm 0.04$ & 1 & $-$ & $0.02 \pm 0.01$ & 2 \\
UBC 577 star \#1 & $-$ & 0 & $0.12 \pm 0.06$ & $0.21 \pm 0.03$ & 2 \\
UBC 577 star \#2 & $0.11 \pm 0.04$ & 1 & $0.02 \pm 0.06$ & $0.04 \pm 0.04$ & 2 \\
UBC 577 star \#3 & $0.10 \pm 0.04$ & 1 & $0.02 \pm 0.06$ & $-0.05 \pm 0.01$ & 2 \\
UBC 577 star \#5 & $0.19 \pm 0.04$ & 1 & $-$ & $0.07 \pm 0.01$ & 2 \\
UBC 60 star \#3 & $0.06 \pm 0.01$ & 2 & $0.15 \pm 0.04$ & $-0.04 \pm 0.01$ & 2 \\
UPK 219 star \#1 & $0.11 \pm 0.02$ & 2 & $0.11 \pm 0.04$ & $0.03 \pm 0.01$ & 2 \\
UPK 84 star \#1 & $0.11 \pm 0.05$ & 1 & $0.22 \pm 0.03$ & $0.04 \pm 0.01$ & 2 \\
UPK 84 star \#2 & $0.20 \pm 0.05$ & 1 & $0.17 \pm 0.03$ & $0.07 \pm 0.01$ & 2 \\
UPK 84 star \#3 & $0.11 \pm 0.01$ & 2 & $0.23 \pm 0.03$ & $0.04 \pm 0.01$ & 2 \\
\end{longtable}

%\FloatBarrier 
%\twocolumn

%\onecolumn

\section{Abundance sensitivities} \label{table:sensitivities}

\begin{longtable}{lcccccc}
\caption{Mo abundance sensitivities to change in stellar parameters. Some sensitivities are not reported either because there was no [Mo/Fe] estimate or because the resulting fit for the test parameters was unsuccessful. }\\
\label{table:sensi_mo} \\
\hline\hline
Star & $\Delta \text{T}{_\text{eff}}$ & $\Delta \log g$  & $\Delta v_{\text{mic}}$ & $\Delta$ [Fe/H] & $\Delta$ [C/Fe] & $\Delta$ [N/Fe]\\
 & (+100 K) & (+0.25 dex) & (+0.5 kms$^{-1}$) &(+0.25 dex) & (+0.25 dex) & (+0.25 dex) \\
\hline
\endfirsthead
\caption{continued.}\\
\hline
Star & $\Delta \text{T}{_\text{eff}}$ & $\Delta \log g$  & $\Delta v_{\text{mic}}$ & $\Delta$ [Fe/H] & $\Delta$ [C/Fe] & $\Delta$ [N/Fe]\\
 & (+100 K) & (+0.25 dex) & (+0.5 kms$^{-1}$) &(+0.25 dex) & (+0.25 dex) & (+0.25 dex) \\
\hline
\endhead
\hline
\endfoot
Alessi 1 star \#2 & 0.14 & 0.00 & 0.00 & -0.29 & -0.07 & 0.00 \\
Alessi 1 star \#3 & 0.13 & 0.01 & -0.01 & -0.28 & -0.06 & 0.00 \\
Alessi 1 star \#5 & 0.14 & 0.01 & 0.00 & -0.32 & -0.16 & -0.08 \\
Alessi 1 star \#6 & 0.14 & 0.00 & 0.00 & -0.28 & -0.05 & 0.00 \\
Alessi 161 star \#1 & 0.15 & 0.02 & 0.00 & -0.31 & -0.15 & -0.07 \\
Alessiteutsch 11 star \#1 & 0.14 & 0.03 & -0.01 & -0.28 & -0.10 & -0.03 \\
Basel 11 star \#1 & 0.14 & 0.01 & 0.00 & -0.33 & -0.14 & 0.00 \\
Basel 11 star \#2 & 0.14 & 0.01 & 0.00 & -0.32 & -0.12 & 0.00 \\
Basel 11 star \#3 & 0.14 & 0.02 & -0.01 & -0.28 & -0.09 & 0.01 \\
COIN-Gaia30 star \#1 & 0.14 & 0.01 & -0.01 & -0.30 & -0.10 & 0.00 \\
Collinder 350 star \#2 & 0.13 & 0.00 & 0.00 & -0.38 & -0.15 & -0.01 \\
Collinder 463 star \#1 & 0.15 & 0.02 & -0.01 & -0.29 & -0.15 & -0.05 \\
Collinder 463 star \#3 & 0.15 & 0.02 & 0.00 & -0.30 & -0.14 & -0.05 \\
Gulliver 51 star \#1 & 0.14 & 0.02 & 0.00 & -0.30 & -0.16 & -0.07 \\
IC 4756 star \#0 & 0.13 & 0.01 & 0.00 & -0.30 & -0.08 & 0.00 \\
IC 4756 star \#2 & 0.14 & 0.03 & -0.02 & -0.27 & -0.08 & -0.03 \\
IC 4756 star \#4 & 0.15 & 0.01 & -0.03 & -0.26 & $-$ & 0.00 \\
IC 4756 star \#5 & 0.14 & 0.00 & 0.01 & -0.37 & -0.28 & -0.07 \\
IC 4756 star \#7 & 0.13 & 0.00 & 0.00 & -0.29 & -0.06 & 0.00 \\
IC 4756 star \#8 & 0.14 & 0.01 & 0.00 & -0.33 & -0.26 & -0.08 \\
IC 4756 star \#9 & 0.13 & -0.01 & 0.00 & -0.30 & -0.06 & 0.00 \\
IC 4756 star \#11 & 0.14 & 0.01 & 0.00 & -0.32 & -0.14 & -0.09 \\
IC 4756 star \#12 & 0.13 & 0.00 & -0.01 & -0.30 & -0.07 & 0.00 \\
IC 4756 star \#13 & 0.13 & -0.01 & 0.00 & -0.31 & $-$ & 0.00 \\
IC 4756 star \#14 & 0.13 & -0.01 & 0.00 & -0.28 & $-$ & 0.00 \\
IC 4756 star \#15 & 0.13 & 0.00 & 0.00 & -0.29 & $-$ & 0.00 \\
LP 1800 star \#1 & 0.15 & 0.02 & -0.01 & -0.36 & -0.13 & -0.12 \\
LP 1800 star \#2 & 0.14 & 0.01 & -0.01 & -0.36 & -0.18 & -0.01 \\
LP 1800 star \#3 & 0.15 & 0.01 & 0.00 & -0.35 & -0.23 & -0.11 \\
NGC 2437 star \#1 & 0.15 & 0.01 & 0.00 & -0.31 & -0.15 & -0.06 \\
NGC 2437 star \#3 & 0.15 & 0.02 & -0.01 & -0.43 & -0.16 & 0.00 \\
NGC 2437 star \#4 & 0.15 & 0.01 & 0.01 & -0.30 & -0.12 & -0.06 \\
NGC 2437 star \#5 & 0.14 & 0.00 & -0.01 & -0.27 & -0.04 & 0.00 \\
NGC 2437 star \#10 & 0.16 & 0.03 & -0.02 & -0.29 & -0.15 & -0.05 \\
NGC 2509 star \#18 & 0.14 & 0.01 & -0.03 & -0.27 & -0.08 & 0.00 \\
NGC 2548 star \#1 & 0.13 & 0.00 & -0.01 & -0.42 & -0.25 & -0.01 \\
NGC 2548 star \#2 & 0.14 & 0.01 & 0.00 & -0.32 & -0.13 & -0.06 \\
NGC 2548 star \#3 & 0.14 & 0.02 & 0.00 & -0.30 & -0.15 & -0.07 \\
NGC 2632 star \#1 & 0.14 & 0.03 & 0.00 & -0.34 & -0.29 & -0.08 \\
NGC2632 hd73665 & 0.15 & 0.02 & 0.01 & -0.33 & -0.27 & -0.08 \\
NGC 2682 star \#8 & 0.14 & 0.02 & -0.01 & -0.29 & -0.12 & -0.05 \\
NGC 2682 star \#9 & 0.14 & 0.02 & -0.01 & -0.30 & -0.15 & -0.08 \\
NGC 6800 star \#1 & 0.13 & 0.01 & -0.01 & -0.29 & -0.07 & 0.00 \\
NGC 6991 star \#4 & 0.13 & 0.00 & 0.00 & -0.28 & -0.05 & 0.00 \\
NGC 6991 star \#6 & 0.20 & 0.00 & -0.01 & -0.32 & -0.16 & -0.08 \\
NGC 6991 star \#7 & 0.13 & 0.00 & -0.01 & -0.27 & -0.04 & 0.00 \\
NGC 6991 star \#9 & 0.13 & 0.00 & 0.00 & -0.27 & -0.04 & 0.00 \\
NGC 6991 star \#10 & 0.14 & 0.00 & 0.01 & -0.32 & -0.14 & 0.06 \\
NGC 7086 star \#1 & 0.15 & 0.03 & -0.01 & -0.29 & -0.12 & -0.05 \\
NGC 7086 star \#2 & 0.15 & 0.03 & -0.01 & -0.28 & -0.12 & -0.05 \\
NGC 7209 star \#2 & 0.16 & 0.02 & 0.00 & -0.36 & -0.50 & -0.12 \\
NGC 752 star \#3 & 0.14 & 0.02 & 0.04 & -0.33 & -0.22 & -0.13 \\
NGC 752 star \#4 & 0.14 & 0.02 & 0.01 & -0.31 & -0.18 & -0.11 \\
NGC 752 star \#5 & 0.14 & 0.02 & 0.02 & -0.32 & -0.19 & -0.11 \\
Ruprecht 171 star \#4 & 0.07 & -0.06 & -0.08 & -0.34 & -0.12 & -0.08 \\
Ruprecht 171 star \#5 & 0.14 & 0.01 & -0.01 & -0.27 & -0.06 & -0.01 \\
Ruprecht 171 star \#6 & 0.07 & -0.07 & -0.08 & -0.34 & -0.12 & -0.08 \\
Ruprecht 171 star \#8 & 0.14 & 0.01 & -0.02 & -0.26 & -0.04 & 0.00 \\
Stock 2 star \#1 & 0.15 & 0.01 & 0.00 & -0.33 & -0.27 & -0.08 \\
Stock 2 star \#2 & 0.14 & 0.03 & -0.02 & -0.23 & -0.05 & -0.03 \\
Stock 2 star \#4 & 0.15 & 0.01 & 0.01 & -0.35 & -0.23 & -0.07 \\
Stock 2 star \#6 & 0.15 & 0.02 & 0.00 & -0.36 & $-$ & -0.09 \\
Stock 2 star \#7 & 0.15 & 0.02 & 0.01 & -0.41 & -0.38 & -0.10 \\
Stock 2 star \#8 & 0.15 & 0.01 & 0.00 & -0.37 & -0.26 & -0.12 \\
Stock 2 star \#9 & 0.14 & 0.01 & 0.00 & -0.35 & -0.24 & -0.07 \\
Stock 2 star \#10 & 0.14 & 0.02 & -0.01 & -0.31 & -0.13 & -0.05 \\
Tombaugh 5 star \#2 & 0.14 & 0.01 & 0.00 & -0.28 & -0.08 & 0.00 \\
Tombaugh 5 star \#3 & 0.14 & -0.02 & 0.01 & -0.32 & -0.12 & 0.01 \\
Tombaugh 5 star \#4 & 0.15 & 0.02 & 0.00 & -0.32 & -0.14 & 0.00 \\
UBC 141 star \#1 & 0.14 & 0.01 & -0.02 & -0.26 & -0.03 & 0.00 \\
UBC 169 star \#2 & 0.14 & 0.01 & -0.01 & -0.35 & -0.13 & -0.01 \\
UBC 194 star \#1 & 0.14 & 0.02 & 0.00 & -0.41 & -0.20 & -0.01 \\
UBC 577 star \#1 & $-$ & $-$ & $-$ & $-$ & $-$ & $-$ \\
UBC 577 star \#2 & 0.13 & 0.00 & 0.00 & -0.31 & 0.02 & 0.00 \\
UBC 577 star \#3 & 0.13 & 0.00 & 0.00 & -0.30 & -0.07 & 0.00 \\
UBC 577 star \#5 & 0.13 & -0.01 & 0.00 & -0.28 & -0.06 & 0.00 \\
UBC 60 star \#3 & 0.14 & 0.01 & -0.01 & -0.31 & -0.26 & -0.07 \\
UPK 219 star \#1 & 0.19 & 0.01 & 0.02 & -0.31 & -0.16 & -0.07 \\
UPK 84 star \#1 & 0.13 & 0.00 & 0.00 & -0.29 & -0.05 & 0.00 \\
UPK 84 star \#2 & 0.13 & 0.00 & 0.00 & -0.28 & -0.04 & 0.00 \\
UPK 84 star \#3 & 0.15 & 0.01 & 0.01 & -0.37 & -0.24 & -0.13 \\
\end{longtable}

\begin{longtable}{lcccccc}
\caption{Ru abundance sensitivities to change in stellar parameters. Some sensitivities are not reported either because there was no [Ru/Fe] estimate or because the resulting fit for the test parameters was unsuccessful. }\\
\label{table:sensi_ru} \\
\hline\hline
Star & $\Delta \text{T}{_\text{eff}}$ & $\Delta \log g$  & $\Delta v_{\text{mic}}$ & $\Delta$ [Fe/H] & $\Delta$ [C/Fe] & $\Delta$ [N/Fe]\\
 & (+100 K) & (+0.25 dex) & (+0.5 kms$^{-1}$) &(+0.25 dex) & (+0.25 dex) & (+0.25 dex) \\
\hline
\endfirsthead
\caption{continued.}\\
\hline
Star & $\Delta \text{T}{_\text{eff}}$ & $\Delta \log g$  & $\Delta v_{\text{mic}}$ & $\Delta$ [Fe/H] & $\Delta$ [C/Fe] & $\Delta$ [N/Fe]\\
 & (+100 K) & (+0.25 dex) & (+0.5 kms$^{-1}$) &(+0.25 dex) & (+0.25 dex) & (+0.25 dex) \\
\hline
\endhead
\hline
\endfoot
Alessi 1 star \#2 & 0.02 & 0.02 & 0.01 & -0.30 & -0.04 & -0.01 \\
Alessi 1 star \#3 & 0.04 & 0.01 & 0.03 & -0.30 & -0.03 & 0.01 \\
Alessi 1 star \#5 & 0.01 & 0.01 & 0.00 & -0.31 & -0.04 & -0.01 \\
Alessi 1 star \#6 & 0.02 & 0.01 & 0.01 & -0.30 & -0.05 & -0.01 \\
Alessi 161 star \#1 & 0.02 & 0.02 & 0.00 & -0.26 & -0.03 & -0.01 \\
Alessiteutsch 11 star \#1 & 0.04 & 0.04 & -0.01 & -0.24 & -0.03 & -0.01 \\
Basel 11 star \#1 & $-$ & $-$ & $-$ & $-$ & $-$ & $-$ \\
Basel 11 star \#2 & $-$ & $-$ & $-$ & $-$ & $-$ & $-$ \\
Basel 11 star \#3 & $-$ & $-$ & $-$ & $-$ & $-$ & $-$ \\
COIN-Gaia30 star \#1 & 0.02 & 0.02 & 0.00 & -0.29 & -0.05 & -0.01 \\
Collinder 350 star \#2 & $-$ & $-$ & $-$ & $-$ & $-$ & $-$ \\
Collinder 463 star \#1 & 0.03 & 0.03 & 0.00 & -0.24 & -0.03 & -0.01 \\
Collinder 463 star \#3 & 0.03 & 0.03 & 0.00 & -0.25 & -0.03 & -0.01 \\
Gulliver 51 star \#1 & 0.03 & 0.03 & 0.00 & -0.25 & -0.04 & -0.01 \\
IC 4756 star \#0 & 0.01 & 0.01 & 0.01 & -0.36 & -0.07 & -0.01 \\
IC 4756 star \#2 & 0.04 & 0.04 & -0.01 & -0.23 & -0.02 & -0.01 \\
IC 4756 star \#4 & 0.04 & 0.04 & -0.01 & -0.24 & -0.03 & -0.01 \\
IC 4756 star \#5 & 0.00 & 0.00 & 0.01 & -0.31 & -0.04 & -0.01 \\
IC 4756 star \#7 & 0.01 & 0.01 & 0.01 & -0.32 & -0.05 & -0.01 \\
IC 4756 star \#8 & 0.01 & 0.01 & 0.01 & -0.33 & -0.04 & -0.01 \\
IC 4756 star \#9 & $-$ & $-$ & $-$ & $-$ & $-$ & $-$ \\
IC 4756 star \#11 & 0.01 & 0.01 & 0.01 & -0.31 & -0.04 & -0.01 \\
IC 4756 star \#12 & 0.01 & 0.01 & 0.00 & -0.33 & -0.05 & -0.01 \\
IC 4756 star \#13 & 0.02 & 0.01 & 0.02 & -0.37 & -0.04 & 0.00 \\
IC 4756 star \#14 & 0.00 & 0.00 & 0.01 & -0.34 & -0.02 & -0.01 \\
IC 4756 star \#15 & 0.02 & 0.02 & 0.00 & -0.30 & -0.04 & -0.01 \\
LP 1800 star \#1 & 0.02 & 0.02 & 0.00 & -0.29 & -0.05 & -0.02 \\
LP 1800 star \#2 & $-$ & $-$ & $-$ & $-$ & $-$ & $-$ \\
LP 1800 star \#3 & 0.01 & 0.01 & 0.01 & -0.31 & -0.05 & -0.02 \\
NGC 2437 star \#1 & -0.03 & 0.02 & 0.00 & -0.31 & -0.04 & -0.01 \\
NGC 2437 star \#3 & $-$ & $-$ & $-$ & $-$ & $-$ & $-$ \\
NGC 2437 star \#4 & 0.02 & 0.02 & 0.00 & -0.27 & -0.04 & -0.01 \\
NGC 2437 star \#5 & -0.01 & 0.00 & -0.03 & -0.32 & -0.06 & -0.03 \\
NGC 2437 star \#10 & 0.04 & 0.04 & -0.02 & -0.25 & -0.03 & -0.01 \\
NGC 2509 star \#18 & $-$ & $-$ & $-$ & $-$ & $-$ & $-$ \\
NGC 2548 star \#1 & $-$ & $-$ & $-$ & $-$ & $-$ & $-$ \\
NGC 2548 star \#2 & 0.00 & 0.04 & -0.01 & -0.33 & -0.03 & -0.03 \\
NGC 2548 star \#3 & 0.03 & 0.03 & 0.01 & -0.25 & -0.03 & 0.00 \\
NGC 2632 star \#1 & 0.02 & 0.02 & 0.00 & -0.28 & -0.03 & -0.01 \\
NGC2632 hd73665 & 0.02 & 0.02 & 0.01 & -0.27 & -0.03 & -0.02 \\
NGC 2682 star \#8 & 0.03 & 0.03 & -0.02 & -0.25 & -0.03 & -0.01 \\
NGC 2682 star \#9 & 0.03 & 0.03 & 0.00 & -0.25 & -0.03 & -0.01 \\
NGC 6800 star \#1 & 0.02 & 0.02 & 0.00 & -0.30 & -0.04 & -0.01 \\
NGC 6991 star \#4 & 0.00 & 0.00 & 0.00 & -0.31 & -0.03 & 0.00 \\
NGC 6991 star \#6 & 0.02 & 0.02 & 0.00 & -0.28 & -0.04 & -0.01 \\
NGC 6991 star \#7 & 0.01 & 0.01 & 0.01 & -0.31 & -0.04 & -0.01 \\
NGC 6991 star \#9 & 0.01 & 0.01 & 0.01 & -0.30 & -0.03 & -0.01 \\
NGC 6991 star \#10 & 0.00 & 0.00 & 0.01 & -0.31 & -0.04 & -0.01 \\
NGC 7086 star \#1 & 0.03 & 0.03 & 0.00 & -0.23 & -0.03 & -0.01 \\
NGC 7086 star \#2 & 0.04 & 0.05 & -0.01 & -0.24 & -0.02 & -0.02 \\
NGC 7209 star \#2 & 0.01 & 0.01 & 0.00 & -0.30 & -0.05 & -0.01 \\
NGC 752 star \#3 & 0.02 & 0.02 & 0.00 & -0.28 & -0.03 & -0.01 \\
NGC 752 star \#4 & 0.02 & 0.02 & 0.00 & -0.28 & -0.03 & -0.01 \\
NGC 752 star \#5 & 0.02 & 0.02 & 0.00 & -0.28 & -0.03 & -0.01 \\
Ruprecht 171 star \#4 & 0.03 & 0.03 & 0.00 & -0.25 & -0.04 & -0.01 \\
Ruprecht 171 star \#5 & 0.03 & 0.03 & 0.00 & -0.25 & -0.04 & -0.01 \\
Ruprecht 171 star \#6 & 0.04 & 0.03 & 0.01 & -0.24 & -0.04 & 0.00 \\
Ruprecht 171 star \#8 & 0.03 & 0.03 & 0.00 & -0.24 & -0.04 & -0.01 \\
Stock 2 star \#1 & 0.01 & 0.01 & 0.01 & -0.30 & -0.04 & -0.01 \\
Stock 2 star \#2 & 0.04 & 0.04 & -0.01 & -0.23 & -0.02 & 0.00 \\
Stock 2 star \#4 & 0.01 & 0.01 & 0.01 & -0.30 & -0.04 & -0.01 \\
Stock 2 star \#6 & $-$ & $-$ & $-$ & $-$ & $-$ & $-$ \\
Stock 2 star \#7 & 0.01 & 0.01 & 0.00 & -0.28 & -0.04 & -0.01 \\
Stock 2 star \#8 & $-$ & $-$ & $-$ & $-$ & $-$ & $-$ \\
Stock 2 star \#9 & 0.02 & 0.02 & 0.00 & -0.34 & -0.05 & -0.01 \\
Stock 2 star \#10 & 0.02 & 0.02 & -0.01 & -0.28 & -0.04 & -0.01 \\
Tombaugh 5 star \#2 & $-$ & $-$ & $-$ & $-$ & $-$ & $-$ \\
Tombaugh 5 star \#3 & $-$ & $-$ & $-$ & $-$ & $-$ & $-$ \\
Tombaugh 5 star \#4 & 0.03 & 0.03 & 0.00 & -0.28 & -0.04 & -0.01 \\
UBC 141 star \#1 & 0.04 & 0.03 & -0.01 & -0.26 & -0.03 & -0.01 \\
UBC 169 star \#2 & $-$ & $-$ & $-$ & $-$ & $-$ & $-$ \\
UBC 194 star \#1 & $-$ & $-$ & $-$ & $-$ & $-$ & $-$ \\
UBC 577 star \#1 & -0.01 & -0.01 & 0.02 & -0.37 & -0.06 & -0.01 \\
UBC 577 star \#2 & 0.00 & 0.00 & 0.01 & -0.34 & -0.14 & -0.01 \\
UBC 577 star \#3 & 0.01 & 0.01 & 0.01 & -0.38 & -0.09 & -0.02 \\
UBC 577 star \#5 & $-$ & $-$ & $-$ & $-$ & $-$ & $-$ \\
UBC 60 star \#3 & 0.02 & 0.02 & 0.00 & -0.29 & -0.03 & -0.01 \\
UPK 219 star \#1 & 0.01 & 0.01 & 0.00 & -0.30 & -0.04 & -0.01 \\
UPK 84 star \#1 & 0.00 & 0.00 & 0.00 & -0.31 & -0.04 & -0.01 \\
UPK 84 star \#2 & 0.00 & 0.00 & 0.01 & -0.33 & -0.05 & -0.01 \\
UPK 84 star \#3 & 0.00 & 0.00 & 0.00 & -0.32 & -0.04 & -0.01 \\
\end{longtable}

\begin{longtable}{lcccccc}
\caption{Zr abundance sensitivities to change in stellar parameters. }\\
\label{table:sensi_zr} \\
\hline\hline
Star & $\Delta \text{T}{_\text{eff}}$ & $\Delta \log g$  & $\Delta v_{\text{mic}}$ & $\Delta$ [Fe/H] & $\Delta$ [C/Fe] & $\Delta$ [N/Fe]\\
 & (+100 K) & (+0.25 dex) & (+0.5 kms$^{-1}$) &(+0.25 dex) & (+0.25 dex) & (+0.25 dex) \\
\hline
\endfirsthead
\caption{continued.}\\
\hline
Star & $\Delta \text{T}{_\text{eff}}$ & $\Delta \log g$  & $\Delta v_{\text{mic}}$ & $\Delta$ [Fe/H] & $\Delta$ [C/Fe] & $\Delta$ [N/Fe]\\
 & (+100 K) & (+0.25 dex) & (+0.5 kms$^{-1}$) &(+0.25 dex) & (+0.25 dex) & (+0.25 dex) \\
\hline
\endhead
\hline
\endfoot
Alessi 1 star \#2 & 0.16 & 0.01 & 0.00 & -0.29 & -0.02 & -0.01 \\
Alessi 1 star \#3 & 0.16 & 0.01 & 0.00 & -0.29 & -0.02 & -0.01 \\
Alessi 1 star \#5 & 0.17 & 0.00 & 0.00 & -0.29 & -0.02 & -0.01 \\
Alessi 1 star \#6 & 0.17 & 0.01 & 0.00 & -0.29 & -0.01 & -0.01 \\
Alessi 161 star \#1 & 0.19 & 0.01 & -0.01 & -0.28 & -0.01 & -0.01 \\
Alessiteutsch 11 star \#1 & 0.20 & 0.02 & -0.03 & -0.27 & -0.01 & -0.01 \\
Basel 11 star \#1 & 0.17 & 0.00 & 0.00 & -0.29 & -0.02 & -0.01 \\
Basel 11 star \#2 & 0.16 & 0.00 & 0.00 & -0.28 & -0.01 & -0.01 \\
Basel 11 star \#3 & 0.19 & 0.01 & -0.02 & -0.28 & -0.01 & -0.01 \\
COIN-Gaia30 star \#1 & 0.18 & 0.01 & -0.01 & -0.28 & -0.01 & -0.01 \\
Collinder 350 star \#2 & 0.15 & 0.00 & 0.00 & -0.31 & -0.02 & -0.02 \\
Collinder 463 star \#1 & 0.20 & 0.02 & -0.02 & -0.27 & -0.01 & -0.01 \\
Collinder 463 star \#3 & 0.20 & 0.01 & -0.01 & -0.28 & -0.01 & -0.01 \\
Gulliver 51 star \#1 & 0.20 & 0.01 & -0.01 & -0.27 & -0.01 & -0.01 \\
IC 4756 star \#0 & 0.16 & 0.01 & 0.00 & -0.29 & -0.01 & -0.01 \\
IC 4756 star \#2 & 0.20 & 0.03 & -0.06 & -0.27 & -0.01 & 0.00 \\
IC 4756 star \#4 & 0.20 & 0.02 & -0.04 & -0.27 & -0.01 & -0.01 \\
IC 4756 star \#5 & 0.16 & 0.00 & 0.00 & -0.28 & -0.01 & -0.01 \\
IC 4756 star \#7 & 0.15 & 0.00 & -0.01 & -0.30 & -0.03 & -0.03 \\
IC 4756 star \#8 & 0.16 & 0.01 & 0.01 & -0.28 & -0.01 & -0.01 \\
IC 4756 star \#9 & 0.15 & 0.00 & 0.01 & -0.29 & -0.02 & -0.01 \\
IC 4756 star \#11 & 0.16 & 0.01 & 0.00 & -0.29 & -0.01 & -0.01 \\
IC 4756 star \#12 & 0.16 & 0.01 & 0.00 & -0.29 & -0.01 & -0.01 \\
IC 4756 star \#13 & 0.15 & 0.00 & 0.01 & -0.29 & -0.01 & -0.01 \\
IC 4756 star \#14 & 0.15 & 0.00 & 0.01 & -0.28 & -0.01 & -0.01 \\
IC 4756 star \#15 & 0.17 & 0.01 & 0.00 & -0.28 & -0.01 & -0.01 \\
LP 1800 star \#1 & 0.18 & 0.00 & -0.01 & -0.30 & -0.02 & -0.02 \\
LP 1800 star \#2 & 0.17 & 0.01 & -0.01 & -0.31 & -0.03 & -0.02 \\
LP 1800 star \#3 & 0.18 & 0.00 & 0.00 & -0.29 & -0.02 & -0.02 \\
NGC 2437 star \#1 & 0.17 & 0.00 & -0.01 & -0.29 & -0.01 & -0.01 \\
NGC 2437 star \#3 & 0.17 & 0.01 & 0.01 & -0.29 & -0.01 & 0.00 \\
NGC 2437 star \#4 & 0.18 & 0.00 & 0.00 & -0.28 & -0.01 & -0.01 \\
NGC 2437 star \#5 & 0.17 & 0.00 & 0.00 & -0.28 & -0.01 & -0.01 \\
NGC 2437 star \#10 & 0.21 & 0.03 & -0.05 & -0.28 & -0.01 & -0.01 \\
NGC 2509 star \#18 & 0.18 & 0.02 & -0.03 & -0.29 & -0.01 & 0.00 \\
NGC 2548 star \#1 & 0.16 & 0.00 & 0.00 & -0.31 & -0.02 & -0.01 \\
NGC 2548 star \#2 & 0.16 & 0.01 & 0.00 & -0.29 & -0.01 & -0.01 \\
NGC 2548 star \#3 & 0.19 & 0.01 & -0.01 & -0.28 & -0.01 & -0.01 \\
NGC 2632 star \#1 & 0.17 & 0.01 & -0.01 & -0.28 & -0.02 & -0.01 \\
NGC2632 hd73665 & 0.18 & 0.01 & 0.00 & -0.29 & -0.02 & -0.01 \\
NGC 2682 star \#8 & 0.19 & 0.01 & -0.02 & -0.28 & -0.01 & -0.01 \\
NGC 2682 star \#9 & 0.19 & 0.01 & -0.02 & -0.28 & -0.01 & -0.01 \\
NGC 6800 star \#1 & 0.16 & 0.01 & -0.01 & -0.28 & -0.01 & -0.01 \\
NGC 6991 star \#4 & 0.16 & 0.00 & 0.00 & -0.28 & -0.01 & -0.01 \\
NGC 6991 star \#6 & 0.18 & 0.01 & -0.01 & -0.28 & -0.01 & -0.01 \\
NGC 6991 star \#7 & 0.17 & 0.01 & 0.00 & -0.28 & -0.01 & -0.01 \\
NGC 6991 star \#9 & 0.16 & 0.00 & 0.01 & -0.28 & -0.01 & -0.01 \\
NGC 6991 star \#10 & 0.16 & 0.00 & 0.00 & -0.28 & -0.01 & -0.01 \\
NGC 7086 star \#1 & 0.20 & 0.02 & -0.02 & -0.27 & -0.01 & -0.01 \\
NGC 7086 star \#2 & 0.20 & 0.03 & -0.03 & -0.27 & -0.01 & 0.00 \\
NGC 7209 star \#2 & 0.18 & 0.01 & 0.00 & -0.29 & -0.02 & -0.01 \\
NGC 752 star \#3 & 0.17 & 0.01 & -0.01 & -0.28 & -0.01 & -0.01 \\
NGC 752 star \#4 & 0.17 & 0.01 & -0.01 & -0.28 & -0.01 & -0.01 \\
NGC 752 star \#5 & 0.17 & 0.01 & -0.01 & -0.28 & -0.01 & -0.01 \\
Ruprecht 171 star \#4 & 0.19 & 0.01 & -0.01 & -0.28 & -0.02 & -0.01 \\
Ruprecht 171 star \#5 & 0.19 & 0.01 & -0.01 & -0.28 & -0.01 & -0.01 \\
Ruprecht 171 star \#6 & 0.19 & 0.01 & -0.01 & -0.28 & -0.02 & -0.01 \\
Ruprecht 171 star \#8 & 0.19 & 0.01 & -0.02 & -0.27 & -0.01 & -0.01 \\
Stock 2 star \#1 & 0.18 & 0.00 & 0.00 & -0.29 & -0.02 & -0.01 \\
Stock 2 star \#2 & 0.19 & 0.03 & -0.05 & -0.27 & -0.01 & 0.00 \\
Stock 2 star \#4 & 0.17 & 0.00 & 0.00 & -0.28 & -0.01 & -0.01 \\
Stock 2 star \#6 & 0.16 & 0.01 & 0.00 & -0.29 & -0.01 & -0.01 \\
Stock 2 star \#7 & 0.18 & 0.01 & 0.00 & -0.28 & -0.01 & -0.01 \\
Stock 2 star \#8 & 0.16 & 0.01 & 0.00 & -0.30 & -0.02 & -0.01 \\
Stock 2 star \#9 & 0.16 & 0.01 & -0.01 & -0.29 & -0.01 & -0.01 \\
Stock 2 star \#10 & 0.18 & 0.01 & -0.03 & -0.28 & -0.01 & -0.01 \\
Tombaugh 5 star \#2 & 0.18 & 0.01 & 0.00 & -0.28 & -0.01 & -0.01 \\
Tombaugh 5 star \#3 & 0.18 & 0.01 & 0.01 & -0.28 & -0.01 & -0.01 \\
Tombaugh 5 star \#4 & 0.19 & 0.01 & -0.01 & -0.29 & -0.02 & -0.01 \\
UBC 141 star \#1 & 0.19 & 0.01 & -0.03 & -0.27 & -0.01 & -0.01 \\
UBC 169 star \#2 & 0.17 & 0.00 & -0.01 & -0.30 & -0.02 & -0.01 \\
UBC 194 star \#1 & 0.16 & 0.03 & -0.01 & -0.30 & -0.02 & -0.01 \\
UBC 577 star \#1 & 0.15 & 0.00 & 0.01 & -0.28 & -0.01 & -0.01 \\
UBC 577 star \#2 & 0.16 & 0.00 & 0.01 & -0.28 & -0.01 & -0.01 \\
UBC 577 star \#3 & 0.16 & 0.01 & 0.00 & -0.29 & -0.02 & -0.01 \\
UBC 577 star \#5 & 0.15 & 0.00 & 0.01 & -0.29 & -0.01 & -0.01 \\
UBC 60 star \#3 & 0.17 & 0.01 & -0.01 & -0.29 & -0.01 & -0.01 \\
UPK 219 star \#1 & 0.18 & 0.01 & -0.01 & -0.29 & -0.01 & -0.01 \\
UPK 84 star \#1 & 0.16 & 0.00 & 0.01 & -0.29 & -0.02 & -0.02 \\
UPK 84 star \#2 & 0.16 & 0.00 & 0.00 & -0.29 & -0.02 & -0.01 \\
UPK 84 star \#3 & 0.15 & 0.00 & 0.00 & -0.29 & -0.02 & -0.02 \\

\end{longtable}

\FloatBarrier 
\twocolumn

\end{appendix}
\end{document}